\documentclass{nature}
\usepackage{pdflscape}
\usepackage{longtable}
\usepackage{hyperref}
\makeatletter
\g@addto@macro{\UrlBreaks}{\UrlOrds}
\makeatother
\usepackage{amsmath}
\usepackage{amssymb}
\usepackage{multibib}

\newcommand{\apjs}{Astrophys. J.}
\newcommand{\apj}{Astrophys. J.}
\newcommand{\apjl}{Astrophys. J.}
\newcommand{\mnras}{Mon. Not. R. Astro. Soc.}
\newcommand{\pasa}{Publ. Astron. Soc. Australia}
\newcommand{\nat}{Nature}

\newcommand{\procspie}{Proc. SPIE.}
\usepackage{rotating}
\usepackage{graphicx}
\usepackage{color}

\newcites{main}{References}
\newcites{methods}{References for Methods}

\newcommand{\zfrbC}{0.378}

\newcommand{\mzzfrb}{z_{\rm FRB}}
\newcommand{\zzfrb}{$\mzzfrb$}

\newcommand{\hgC}{J212258.0-792350}  

\newcommand{\Nfrbs}{four}
\newcommand{\listFRBs}{190102, 190608, 190611 and 190711}

\newcommand{\moh}{\Omega_b h_{70}}   
\newcommand{\oh}{$\moh$}
\newcommand{\rhoczero}{\rho_{c,0}}  

\newcommand{\rmunits}{{\rm rad \, m^{-2}}}

\newcommand{\mlike}{\mathcal{L}}
\newcommand{\like}{$\mlike$}

\newcommand{\mdmunits}{{\rm pc \, cm^{-3}}} 
\newcommand{\dmunits}{$\mdmunits$}
\newcommand{\mdmfrb}{{\rm DM}_{\rm FRB}}
\newcommand{\dmfrb}{$\mdmfrb$}
\newcommand{\mdmpfrb}{{\rm DM'}_{\rm FRB}}
\newcommand{\dmpfrb}{$\mdmpfrb$}
\newcommand{\mdmcosmic}{{\rm DM}_{\rm cosmic}}
\newcommand{\dmcosmic}{$\mdmcosmic$}
\newcommand{\mdmacosmic}{\langle {\rm DM}_{\rm cosmic} \rangle}
\newcommand{\dmacosmic}{$\mdmacosmic$}
\newcommand{\mdmism}{{\rm DM}_{\rm MW,ISM}}
\newcommand{\dmism}{$\mdmism$}
\newcommand{\mdmmw}{{\rm DM}_{\rm MW,halo}}
\newcommand{\dmmw}{$\mdmmw$}
\newcommand{\mdmhost}{{\rm DM}_{\rm host}}
\newcommand{\dmhost}{$\mdmhost$}
\newcommand{\mdmnui}{{\rm DM}_{\rm host}}
\newcommand{\dmnui}{$\mdmnui$}
\newcommand{\msdm}{\sigma_{\rm DM}}
\newcommand{\sdm}{$\msdm$}

\newcommand{\mpcosmic}{p_{\rm cosmic}(\Delta)
}
\newcommand{\pcosmic}{$\mpcosmic$}
\newcommand{\mshost}{\sigma_{\rm host}}
\newcommand{\mphost}{p_{\rm host}(\mdmhost|\mu,\mshost)} 
\newcommand{\phost}{$\mphost$}
\newcommand{\mobh}{\Omega_b h_{70}}
\newcommand{\obh}{$\mobh$}

\newcommand{\vobh}{0.056} 
\newcommand{\iobh}{[0.046, 0.066]} 

\newcommand{\figdmcosmic}{2}
\newcommand{\fighostmontage}{1}      
\newcommand{\figfrbB}{2} 
\newcommand{\figfrbC}{3}
\newcommand{\figfrbA}{Extended Data Figure 1}
\newcommand{\fighostDM}{Extended Data Figure 2}
\newcommand{\figMoreModels}{Extended Data Figure 3}
\newcommand{\figsevenburstcorner}{Extended Data Figure 4}
\newcommand{\figmcmccorner}{Extended Data Figure 5}
\newcommand{\figobh}{3}

\newcommand{\tabFRBs}{1}
\newcommand{\tabDMcutoff}{Extended Data Table 1}
\newcommand{\tabMCMC}{Extended Data Table 2}

\def\araa{Annu.\,Rev.\,Astron.\,Astrophys.}             
\def\apj{Astrophys.\,J.}                 
\def\apjl{Astrophys.\,J.}                
\def\apjs{Astrophys.\,J.\,Supp.}               
\def\aap{Astron. Astrophys.}                
\def\mnras{Mon.\,Not.\,R.\,Astron.\,Soc.}             
\def\nat{Nature}              
\def\pasp{Publ.\,Astron.\,Soc.\,Pacific}               
\def\pasa{Publ.\,Astron.\.Soc.\,Aust.}	

\title{A census of baryons in the Universe from localized fast radio bursts}

\author{
J.-P.~Macquart$^{1}$, J.X.~Prochaska$^{2,3}$, M.~McQuinn$^{4}$, K.W.~Bannister$^{5}$, S.~Bhandari$^{5}$, C.K.~Day$^{6,5}$, A.T.~Deller$^{6}$, 
R.D.~Ekers$^{5,1}$, C.W.~James$^{1}$, L.~Marnoch$^{7,5}$, S.~Os{\l}owski$^{6}$, C.~Phillips$^{5}$, S.D.~Ryder$^{7}$, D.R.~Scott$^{1}$, R.M.~Shannon$^{6}$, N.~Tejos$^{8}$
}

\begin{document}

\maketitle

\begin{affiliations}
\item International Centre for Radio Astronomy Research, Curtin Institute of Radio Astronomy, Curtin University, Perth, WA 6845, Australia.
\item University of California Observatories--Lick Observatory, University of California, Santa Cruz, CA 95064, USA. 
\item Kavli Institute for the Physics and Mathematics of the Universe, 5-1-5 Kashiwanoha, Kashiwa 277-8583, Japan.
\item Astronomy Department, University of Washington, Seattle, WA 98195, USA.
\item Commonwealth Science and Industrial Research Organisation, Australia Telescope National Facility, PO Box 76, Epping, NSW 1710, Australia.
\item Centre for Astrophysics and Supercomputing, Swinburne University of Technology, Mail H30, PO Box 218, Hawthorn, VIC 3122, Australia.
\item Department of Physics and Astronomy, Macquarie University, North Ryde, NSW 2109, Australia.
\item Instituto de F\'isica, Pontificia Universidad Cat\'{o}lica de Valpara\'{i}so, Casilla 4059, Valpara\'{i}so, Chile.
\end{affiliations}


\newpage


\begin{abstract}
More than three quarters of the baryonic content of the Universe resides in a highly diffuse state that is difficult to observe, with only a small fraction directly observed in galaxies and galaxy clusters\citemain{FukugitaHoganPeebles98,CenOstriker06}.  Censuses of the nearby Universe have used absorption line spectroscopy\citemain{Shull12,Nicastro18} to observe these invisible baryons, but these measurements rely on large and uncertain corrections and are insensitive to the majority of the volume, and likely mass.  Specifically, quasar spectroscopy is sensitive either to only the very trace amounts of Hydrogen that exists in the atomic state, or highly ionized and enriched gas\citemain{tripp+02,tumlinson+11,Nicastro18} in denser regions near galaxies\citemain{x+11}. Sunyaev-Zel'dovich analyses\citemain{Hojjatietal2017,deGraaffetal2019} provide evidence of some of the gas in filamentary structures and studies of X-ray emission are most sensitive to gas near galaxy clusters\citemain{Eckertetal2015,deGraaffetal2019}.
Here we report the direct measurement of the baryon content of the Universe using the dispersion of a sample of localized fast radio bursts (FRBs), thus utilizing an effect that measures the electron column density along each sight line and  accounts for every ionised baryon\citemain{McQuinn14,Macquart15,Prochaska19}.  
We augment the sample of published arcsecond-localized FRBs\citemain{Chatterjeeetal2017,Bannister19,Prochaskaetal19,Ravi19,Marcote2020} with a further four new localizations to host galaxies which have measured redshifts of $0.291$, $0.118$, $0.378$ and $0.522$, completing a sample sufficiently large to account for dispersion variations along the line of sight and in the host galaxy environment\citemain{McQuinn14} to derive a cosmic baryon density of $\Omega_{b} = 0.051_{-0.025}^{+0.021} \, h_{70}^{-1}$ (95\% confidence).  This independent measurement is consistent with Cosmic Microwave Background and Big Bang Nucleosynthesis values\citemain{cooke18,planck15}.  
\end{abstract}

The Commensal Real-time ASKAP Fast Transients (CRAFT) survey on the Australian Square Kilometre Array Pathfinder (ASKAP) has commissioned a mode capable of localizing fast radio bursts with sub-arcsecond accuracy, thus enabling identification of their host galaxies and measurement of their redshifts.  ASKAP consists of 36 antennas equipped with phased array feeds, able to view 30\,deg$^{2}$ on the sky.
Bursts are detected by incoherently summing the total power signal of individual beams from each of the antennas.  Bursts detected in the incoherent pipeline are subsequently localized interferometrically by triggering a download of voltage data from a 3.1\,s-duration ring buffer that is correlated and imaged at high time resolution to provide the localizations \citemain{Bannister19,Prochaskaetal19}.  The 6\,km baselines of the array yield statistical position errors  $\approx 10^{\prime \prime}\,(S/N)^{-1}$, where the final coherent signal-to-noise of the burst, $S/N$, exceeds $50$ for any burst whose signal-to-noise in the incoherent pipeline is greater than 9. The resulting statistical (thermal) uncertainties are smaller than $0.2^{\prime \prime}$. Systematic errors in these positions are typically smaller than $0.5^{\prime \prime}$.  
At $z=0.5$, $1^{\prime \prime}$ corresponds to 5\,kpc which is approximately the precision needed to associate an FRB to its host galaxy while reducing the chance coincidence probability to $< 1\%$ \citemain{Prochaskaetal19}.

We report the detection of \Nfrbs\  localized ASKAP bursts.  Table \tabFRBs\ lists the burst properties, sky positions and host galaxy offsets, while Figure \fighostmontage\ shows the host galaxy identifications (see also\citemain{Bhandari19} and Methods).  
Their dispersion measures (DMs) are well in excess of the $30-100\,$pc\,cm$^{-3}$ contributions expected of the disk and halo of the Milky Way at high Galactic latitudes\citemain{NE2001,Prochaska19}, with the large excesses attributable to the IGM and gas within each burst host galaxy.  Two other ASKAP-detected bursts and their host galaxies were reported previously\citemain{Bannister19,Prochaskaetal19} in addition to three other host-galaxy identifications\citemain{Chatterjeeetal2017,Ravi19,Marcote2020}.

The precise localization of a set of FRBs
to their host galaxies provides the first ensemble
of \dmfrb\ and \zzfrb\ measurements.
The \dmfrb\ measurement represents the 
electron density weighted by $(1+z)^{-1}$ integrating over all physical distance increments $ds$ to a given FRB: $\mdmfrb = \int n_e \, ds / (1+z)$.
Physically, we expect \dmfrb\ to separate into
four primary components:
\begin{equation}
\mdmfrb(z) = \mdmism + \mdmmw + \mdmcosmic(z) + \mdmhost(z)
\end{equation}
with \dmism\ the contribution from our Galactic ISM,
\dmmw\ the contribution from our Galactic halo\citemain{Prochaska19},
\dmhost\ the contribution from the host galaxy including its
halo and any gas local to the event,
and \dmcosmic\ the contribution from all
other extragalactic gas.
Only \dmcosmic, determined by its path length through the intergalactic medium and the increase in baryon density with look-back time, is expected to have a strong redshift dependence, although \dmhost\ is weighted by $(1+\mzzfrb)^{-1}$ and may correlate with age, e.g.\ if host galaxies have systematically lower mass at earlier times.

Adopting our cosmological paradigm of a flat universe
with matter and dark energy, the average value of \dmcosmic\ 
to redshift \zzfrb\ is: 
\begin{equation}
    \mdmacosmic = \int\limits_0^{\mzzfrb} \frac{c \bar n_e(z) dz}{
    H_0 (1+z)^2 \sqrt{\Omega_m (1+z)^3 + \Omega_\Lambda}}  
    \label{DMzRelation}  
\end{equation}
with  $\bar n_e = f_d \rho_b(z) m_p^{-1} (1-Y_{\rm He}/2)$, where $m_p$ is the proton mass, $Y_{\rm He} = 0.25$ is the mass fraction of Helium, assumed doubly ionized in this gas,
$f_d(z)$ is the fraction of cosmic baryons in diffuse ionized gas
(this accounts for dense baryonic phases, e.g.\ stars, neutral gas;
see Methods), 
$\rho_b(z) = \Omega_b \rhoczero (1+z)^3$,
and $\Omega_m$ and $\Omega_\Lambda$
are the matter and dark energy densities 
today in units of $\rhoczero= 3 H_0^2/8 \pi G$ where we parameterize Hubble's constant $H_0$ in terms of the dimensionless $h_{70} = H_0/(70 \, \rm km \, s^{-1} \, Mpc^{-1})$.
The \dmism\ term arises primarily from
the so-called warm ionized medium (WIM) of the Galaxy
and is estimated from a model of this ISM
component\citemain{NE2001,gaensler08}.
At the high Galactic latitudes ($|b| > 33$\,deg)
of the ASKAP sample, the value is $\mdmism \approx 30 \, \mdmunits$.
The \dmmw\ term 
is not well constrained\citemain{Prochaska19},
but is expected to be in the range $\approx 50-100 \mdmunits$.
Hereafter we assume $\mdmmw = 50 \, \mdmunits$
and emphasize that the sum of its scatter and uncertainty 
are less than those of \dmcosmic\ and \dmhost,
which we discuss below.


%
%
Figure~\figdmcosmic\ shows the theoretical curve for 
\dmacosmic\ vs.\ \zzfrb\ 
for the Planck~15 cosmology\citemain{planck15}
and a model estimate of the scatter (90\%\ interval) due to statistical
variations in foreground cosmic structure (see Methods).
Overplotted on the model 
are the estimated \dmcosmic\ and measured
\zzfrb\ values for all arc-second 
localized FRBs.  We have estimated \dmcosmic\ 
by subtracting from the measured \dmfrb\ value
\dmism\ from the Galactic ISM model, our assumed 
\dmmw\ contribution, and an ansatz of
$\mdmhost = 50/(1+z) \mdmunits$ estimated from theoretical
work and informed from the analysis below.
We ignore FRB~121102 and FRB~190523 in the majority of analysis
that follows due to selection bias in their discovery, FRB~180916 due to its low Galactic latitude (see Methods), and 
FRB~190611 due to its tentative association to a host galaxy (see Methods).
The five ASKAP FRBs that remain comprise what we term the gold-standard
sample.
The agreement between model and data 
is striking.
Effectively, the FRB measurements confirm the presence of baryons
with the density estimated from the CMB and BBN, and these five measurements are consistent with all the missing baryons being present in the ionized intergalactic medium.

This result motivated us to quantitatively test
for consistency of \oh\ with CMB and BBN measurements, simultaneously determining the uncertain host galaxy contributions to \dmfrb\ as well as the sightline-to-sightline variance in dispersion owing to the IGM. 
We do this by analysing the joint likelihood of our sample of five (seven, including FRBs~190523 and 190611) \dmfrb,~\zzfrb\ measurements against a four parameter model: one parameter for the large-scale structure scatter in \dmcosmic, two parameters for \dmhost\ (a mean and a scatter), and \obh. Our model for the contributions to \dmfrb\ starts with equations 1 and 2, and we develop parametric models for \dmhost\ and the intrinsic scatter in \dmcosmic. 
 We again fix $\mdmmw = 50 \, \mdmunits$
 and adopt the Galactic ISM model\citemain{NE2001}
 for \dmism.  Uncertainty in these values can be absorbed into our model for \dmhost. 


For \dmcosmic, our model accounts for scatter
in the electron column from foreground structures, which is largely caused by random variation in the number of halos a given sightline intersects.  Cosmological simulations show that this variation is sensitive to the extent by which galactic feedback redistributes baryons around galactic halos \citemain{McQuinn14,Prochaskaetal19} and that the fractional standard deviation of the cosmic DM equals approximately\citemain{McQuinn14}  $F z^{-1/2}$ for $z<1$, where the parameter $F$ quantifies the strength of the baryon feedback 
(0.1 being strong feedback and 0.4 being weak).  Stronger feedback corresponds to situations in which feedback processes expel baryons to larger radii from their host galaxies or where more massive halos are evacuated by such feedback.  The formalism incorporates the effect of large scale structure associated with voids and the intersection of sightlines with clusters. We find a one-parameter model that assumes a motivated shape for the probability distribution of \dmcosmic\ given $F$ provides a remarkably successful description of a wide range of cosmological simulations (see Methods). 
 Our form for the distribution is strongly asymmetric towards lower redshifts, admitting large \dmcosmic\ values that we find are important in the estimation of \obh.

We chose our model for \dmhost\ to follow a log-normal distribution, characterised by a median $\exp(\mu)$ 
and logarithmic width parameter $\sigma_{\rm host}$ such that the standard deviation of the distribution is $\exp(\mu) e^{\sigma_{\rm host}^2/2} (e^{\sigma_{\rm host}^2} -1)^{1/2}$. We do not attempt to incorporate redshift-dependent evolution in the host galaxy dispersion contribution, however we do scale the distribution of ${\rm DM}_{\rm host}$ by the factor $(1+z_{\rm host})^{-1}$ applicable to a parcel of plasma at redshift $z_{\rm host}$ so that ${\rm DM}_{\rm host}$ is interpreted as the dispersion in the rest frame of the host galaxy.  Our choice of a log-normal distribution is conservative in that it allows for a tail extending to large positive values, which may not be present in our sample given our selection criteria (see Methods) and the burst locations relative to the host stellar surface density. We explore \dmhost\ distributions with median values in the range $\mu=20-200\,\mdmunits$ and $\sigma_{\rm host}$ in the range $0.2-2.0$.

Our final analysis compares the relative likelihood of models in the four parameters \obh$,~ F, \sigma_{\rm host},~\mu$ parameter space (see Methods for a Bayesian approach that yields similar constraints).
Marginalizing the other parameters over motivated ranges,
we derive the constraints on \oh\ shown in Figure~\figfrbC\ using our five-FRB gold sample.  The results are fully consistent with the joint CMB+BBN 
estimations and with only five (seven) burst redshifts the experiment yields
a precision of $\sigma(\moh)/\moh = 0.31$ ($0.28$) at the 68\% confidence level, with $F$ marginalized over the range $[0.09,0.32]$ (see Methods).
This quantitative result on \oh\ substantiates our
inference from the DM-$z$ relation in Figure~\figfrbB\ that
the FRB ensemble has resolved the missing baryons problem.  The ratio of our estimated $\Omega_{b}$ to that from Cosmic Microwave Background and Big Bang Nucleosynthesis measurements is $1.1_{-0.6}^{+0.5} \,h_{70}$.  Formally, we exclude \oh\ $< 0.02$ ($0.01$) at the 98.6\% (99.8\%) confidence level. This constraint should improve considerably in the near term as ASKAP and other facilities acquire a larger sample of bursts with redshifts. 

Additionally, analysis of our gold-standard sample mildly favors a median host galaxy contribution of  $\sim 100 \, \mdmunits$ with a factor of two dispersion around this value ($\sigma_{\rm host} \sim 1$). 
This quantifies our result that the host contributions are sufficiently small to not compromise the use of FRBs for cosmology and intergalactic medium science.  Even with our current small sample we are beginning to constrain viable models for the redistribution of the cosmic baryons by galactic feedback.  If we adopt a prior on \oh\ from the CMB, BBN, and supernovae surveys\citemain{gold}, 
we find $F=0.04_{-0.04}^{+0.26}$ (68\% confidence), and if we further
include FRB\,190523 and FRB\,190611 we find $F=0.23_{-0.12}^{+0.27}$ (see Methods).  
 A factor of two smaller error would start to differentiate between viable feedback scenarios (as discussed further in Methods),
 suggesting that with modestly larger samples, FRBs have not only revealed that all the baryons are present but will constrain where they lie.




\bibliographystylemain{naturemag}


\clearpage

\begin{figure}[h] \label{hostsImages}
\includegraphics[width=0.8\linewidth]{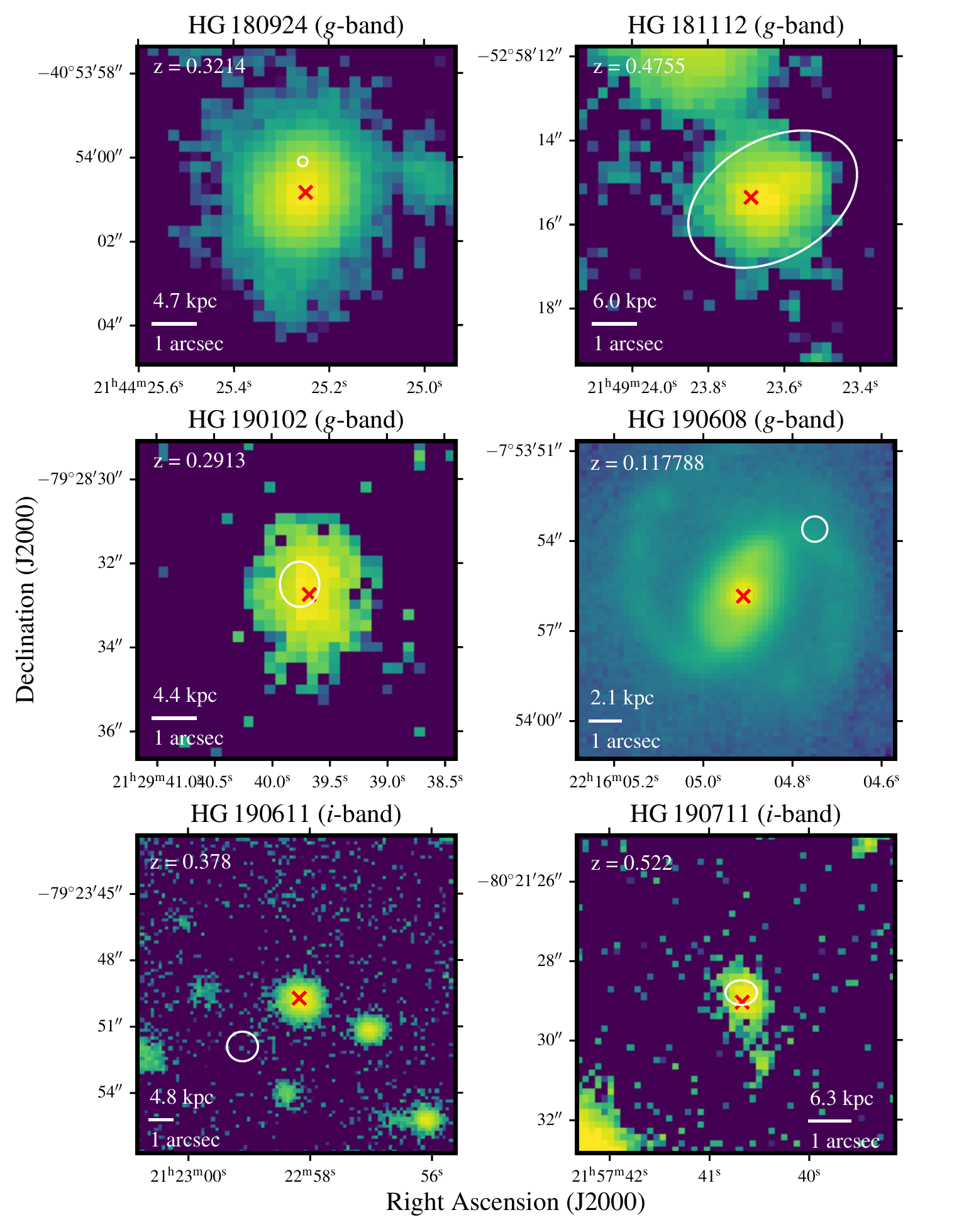}

Figure \fighostmontage: {\bf Locations of FRBs relative to their host galaxies.}  
Deep GMOS (FRBs 190611 \& 190711) and VLT (all other FRBs) optical images of the host galaxies of the bursts localized by ASKAP, including the four new bursts reported here.  White ellipses denote the 90\% confidence region of each burst position, including statistical uncertainty and phase referencing errors, while the the red crosses mark the measured centroids of each host galaxy.  The identification of the host galaxy of FRB\,190611 is tentative.
\end{figure}

\begin{figure}[h]
\includegraphics[width=\linewidth]{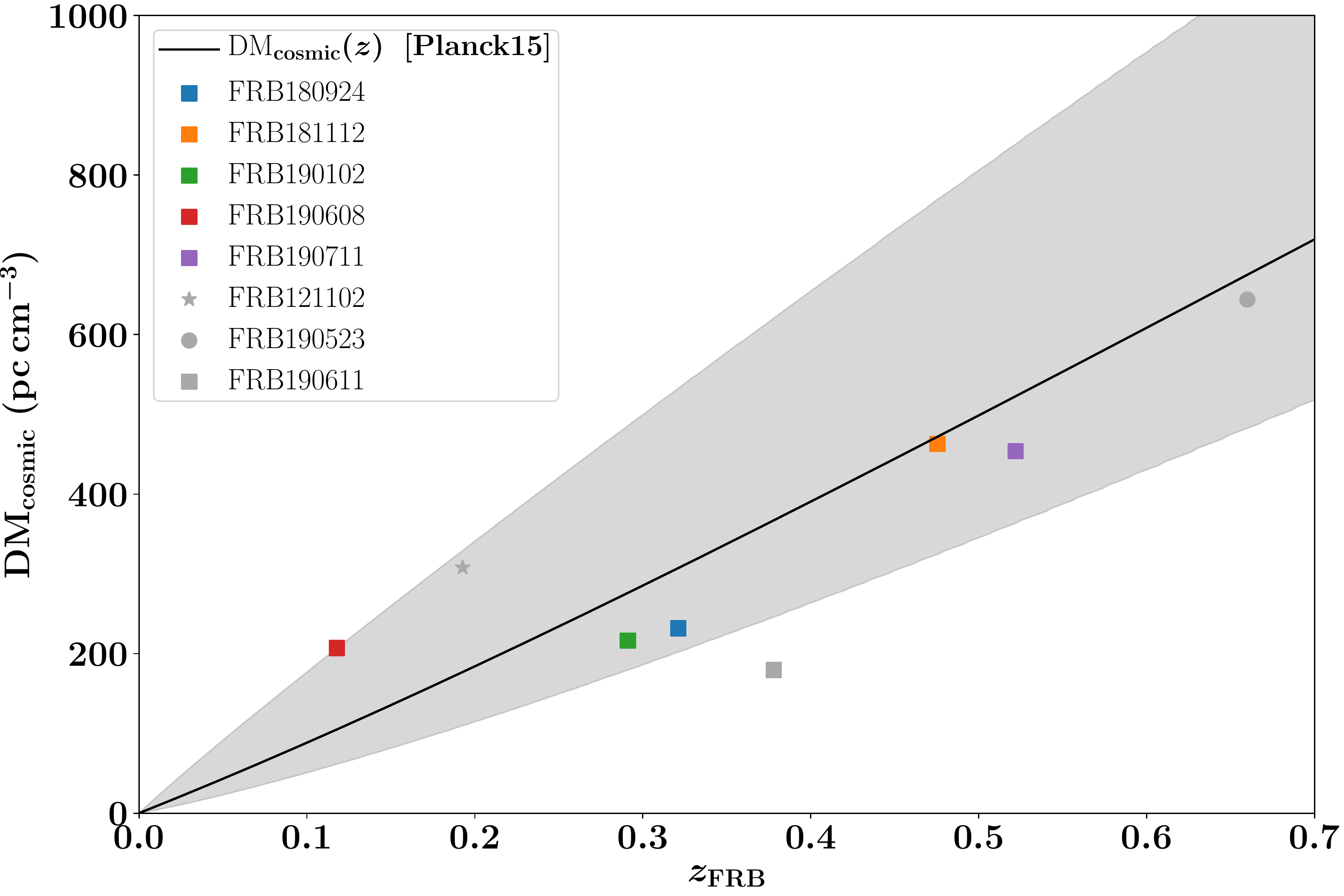} 
    Figure \figfrbB: {\bf The DM-redshift relation for localized FRBs.} Points are estimations of the cosmic 
    dispersion measure
    \dmcosmic\ vs.\ FRB redshift \zzfrb\
    for all current arcsecond- and subarcsecond-localized FRBs.  
    The \dmcosmic\ values are derived by correcting the observed dispersion measure \dmfrb\ for the estimated contributions from our Galaxy and the FRB host galaxy (the latter assumed here to be $50 (1+z)^{-1}\,$pc\,cm$^{-3}$; see text for details).  
    Coloured points represent the `gold standard sample' upon which
    our primary analysis is based.
    The solid line denotes the expected relation between \dmcosmic\ and redshift for a universe based on the Plank15 cosmology (i.e. $\Omega_b=0.0486$ and $H_0=67.74\,$km\,s$^{-1}$\,Mpc$^{-1}$).  
    The shaded region encompasses 90\%\ of the \dmcosmic\ values from a model for ejective feedback in Galactic halos that is motivated by some simulations
    (with $F=0.2$ in Equation~\ref{eqn:MHR} in Methods), illustrating that the observed scatter is largely consistent with the scatter from the intergalactic medium.
    
\end{figure}

\begin{figure}[htbp!]
\includegraphics[width=0.75\linewidth]{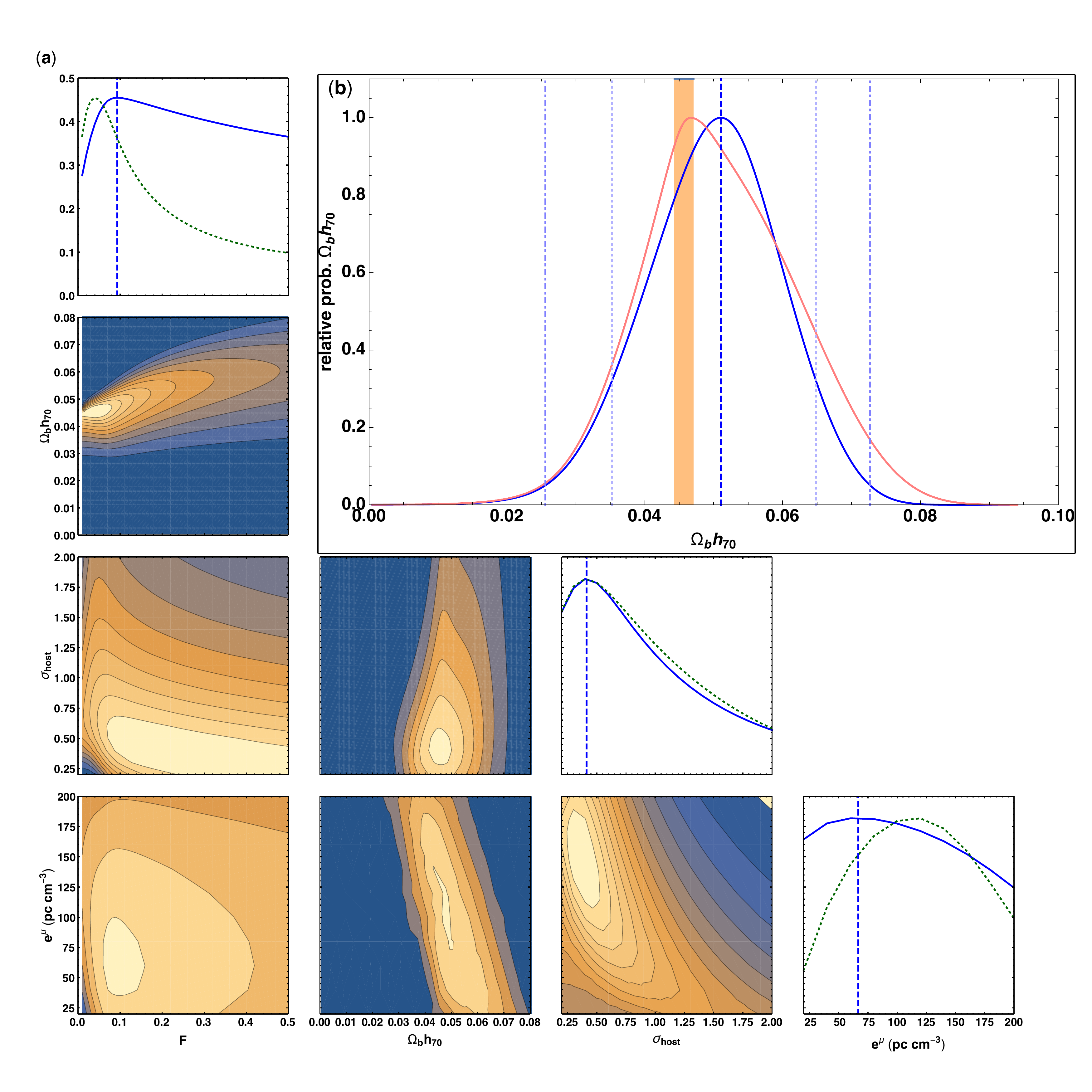} 

Figure \figfrbC: {\bf The density of cosmic baryons derived from the FRB sample.} The constraints on the IGM parameters $\Omega_b h_{70}$ and $F$, and the host galaxy parameters $\mu$ and $\sigma_{\rm host}$ for a log-normal DM distribution are derived using the five gold-standard bursts (as described in the text and Methods). The corner plots in panel (a) display the probability that a given value of $F$, $\mu$ or $\sigma_{\rm host}$ is the consistent with the data against their most likely values, and marginalised over the other parameters: heavy dashed lines represent the most likely values in each case.  The green, dotted lines in the corner plots of $F$, $e^\mu$ and $\sigma_{\rm host}$ denote the relative likelihood of these parameters when $\Omega_b h_{70}$ is constrained to the value set by the CMB+BBN measurements. The contours displayed are in increments of 10\% of the peak value.  
    Panel (b) displays the corner plot for \obh\ where the orange shaded region denotes the range to which $\Omega_b h_{70}$ is confined by CMB+BBN measurements. The dotted and dot-dashed lines represent the 68\% and 95\% confidence intervals of each parameter respectively. The distribution of $\Omega_b$ is alternately marginalised over the range of $F$ indicated by cosmological simulations, $[0.09,0.32]$ (blue curve; see Methods), and over the entire range of F $[0,0.5]$ investigated here (red curve).  
\end{figure}

\clearpage




 

\begin{landscape}
\footnotesize{
\begin{longtable}[c]{lllllllll}
\hline
\hline
  \multicolumn{1}{c}{FRB} & \multicolumn{1}{c}{Time}  & \multicolumn{1}{c}{DM} & \multicolumn{1}{c}{RM} &  \multicolumn{1}{c}{$E_\nu$} & \multicolumn{1}{c}{R.A.}  & \multicolumn{1}{c}{Dec.}   & host redshift & offset from host \\
\null  &  \multicolumn{1}{c}{(UTC$^{(1)}$)} & \multicolumn{1}{c}{(pc\,cm$^{-3}$)} & 
\multicolumn{1}{c}{(rad\,m$^{-2}$)} & \multicolumn{1}{c}{(Jy\,ms)$^{(2)}$ }& \multicolumn{1}{c}{(hh:mm:ss.s)$^{3}$} & \multicolumn{1}{c}{(dd:mm:ss)$^{3}$} & \null & \multicolumn{1}{c}{nucleus (kpc)} \\
\hline
180924 &  16:23:12.6265 & 361.42(6) & 14(1) & 16 $\pm$ 1 & 21:44:25.255 $\pm$ 0.006 $\pm$ 0.008 & -40:54:00.10 $\pm$ 0.07 $\pm$ 0.09   & 0.3214 & $3.5 \pm 0.9$ \\ 
181112 &  17:31:15.48365 & 589.27(3) & 10.9(9) & 26(3)  & 21:49:23.63 $\pm$ 0.05 $\pm$ 0.24  & -52:58:15.4 $\pm$ 0.3 $\pm$ 1.4&  0.4755 & $3.1_{-3.1}^{+15.7}$ \\ 
190102 &  05:38:43.49184 & 363.6(3)  & 110 & 14(1) & 21:29:39.76 $\pm$ 0.06 $\pm$ 0.16 &  -79:28:32.5 $\pm$ 0.2$ \pm$ 0.5  & 0.291 & $1.5_{-1.5}^{+3.4}$ \\ 
190608 &  22:48:12.88391 & 338.7(5) & -- & 26(4) & 22:16:04.75 $\pm$ 0.02 $\pm$ 0.02 & -07:53:53.6 $\pm$ 0.3 $\pm$ 0.3 &  0.1178 & $7.0 \pm 1.3$ \\ 
190611 & 5:45:43.29937 & 321.4(2) & -- & 10(2) & 21:22:58.91 $\pm$ 0.11 $\pm$ 0.23 & -79:23:51.3 $\pm$ 0.3 $\pm$ 0.6 & 0.378 & $17.2 \pm 4.9$ \\ 
190711 & 01:53:41.09338 &  593.1(4) & -- & 34(3) &   21:57:40.68 $\pm$ 0.051 $\pm$ 0.15  & -80:21:28.8 $\pm$ 0.08 $\pm$ 0.3  & 0.522 & $1.5^{+3.6}_{-1.5}$  \\

\hline 
\end{longtable} 
}
\begin{flushleft}
Table 1: {\bf Properties of FRBs interferometrically localised with ASKAP.}   
(1) Burst arrival time referenced to a frequency of $1152$~MHz
(2) Quoted errors on the last significant digit of the fluence represent a 90\% confidence limit.
(3) Errors listed after the burst position represent the statistical and systematic uncertainties respectively, and are combined in quadrature for a final absolute positional uncertainty.
\end{flushleft}
\end{landscape}

\newpage
\begin{methods}


\subsection{Sample Selection}
\label{sec:sample}

Analogous to cosmological studies of the distance ladder
with supernovae, we wish to establish a strict set of criteria
for the FRB sample to minimize biases while maximizing statistical
power.  On the latter point, we wish to construct the largest
sample while avoiding events whose DM is dominated by 
non-cosmological effects (i.e.\ host or Galactic gas).
Regarding bias, the greatest concern is the association
of a host galaxy to a given FRB on the basis of the
DM-$z$ relation, i.e.\ adopting this relation as a prior
to establish the host.  This is a valuable practice when
one aims to resolve the underlying host galaxy 
population\citemain{Ravi19,Mahony18} but would bias any cosmological study.
Last, one must also be cognizant of biases related to triggering
on FRB events.  This practice is a complex function of the 
FRB fluence, its DM, and the pulse properties\citemain{Connor19}.

With these issues in mind, we propose the following set of
criteria to generate a `gold sample' of FRBs for cosmological
study:

\begin{enumerate}
 
    \item To make a confident host galaxy association we require a probability of mis-identifications to be $< 1$\% without invoking the DM-$z$ relation since to do so would introduce a bias. 
    For this we require the 95\% localization area to encompass one and only one galaxy unless multiple galaxies have a common redshift.  By ``encompass'' we include significant light from any part of the galaxy.  In practice, this will require a localization to $< 1''$ for $z>1$, but becoming less stringent for less distant hosts.  We propose an initial set of specific criteria as follows:
    \begin{itemize}
    \item[i]  Define 95\% areas for the localization and for
each galaxy in the region.  Call these $L$ and $G_1$, $G_2$, etc.

\item[ii]  Demand one and only one galaxy overlap $L$.  The only exception is if $z_1 = z_2$.

\item[iii]  For the overlapping $L$ and $G$, require that $>50\%$ of the smaller
area lie within the larger.

\item[iv]  Do this for galaxies as faint as $R=25$ (anything fainter is generally 
too difficult for a spectroscopic redshift anyhow).
    \end{itemize}
    
    \item The finite temporal and spectral resolution of the FRB survey causes a decline in sensitivity with increasing DM to the point that telescope resolution causes an effective threshold at ${\rm DM}_{\rm cutoff}$ at which point a burst would no longer have been detectable.  A conservative approach would omit any burst with ${\rm DM}_{\rm cutoff}$ sufficiently low that it excludes a large ($\gtrsim 30$\%) fraction of the total probability of $p({\rm DM}_{\rm total}|z)$, on the grounds that it presents a biased probe of $p({\rm DM}_{\rm total}|z)$.  Although application of this criterion presupposes a DM-$z$ relation and its probability density function, it does so only weakly.  This point is addressed in detail below in the subsection on biases in the probability distribution.
    
An event detected near to the sensitivity threshold is biased in the sense that the instrumental decline in sensitivity with increasing DM dictates that any burst detected near this threshold would not have been detectable at higher DMs.  Thus, for a given redshift we are biased to finding events with DMs that are under-representative of the entire DM distribution at that redshift.  Thus, only more luminous bursts, whose detection S/N is sufficiently high that ${\rm DM}_{\rm cutoff}$ exceeds the plausible range of ${\rm DM}_{\rm total}$ at that redshift are devoid of this bias.
    
    \item FRB events with extreme properties (e.g.\ high RM,
    large temporal broadening) will be excluded to minimize the
    impact of host galaxy and Galactic gas.  
    
    \item A cutoff is imposed on FRBs whose expected contribution from the disk component of the Milky Way ISM is large, to avoid large uncertainties in the subtraction of the Galactic ISM DM contribution.  Models of the Galactic plasma distribution\citemain{NE2001} typically produce errors in known pulsar distances of order several tens of percent (and much higher in some cases)\citemain{Schnitzeler12,YMW16}.  To avoid DM errors in excess of $\approx 100\,$\dmunits\ we restrict our sample to those bursts whose predicted ${\rm DM}_{\rm ISM}$ values are less than 100\,\dmunits.  A conservative application of this criterion restricts FRB detections to sight lines at high Galactic latitude $|b| \gtrsim 20\,$deg.
    
\end{enumerate}
We acknowledge that these criteria are subject to refinement
as we learn more about FRB progenitors and their host galaxies.

Regarding Criterion 3, a dominant contributor to the DM variance 
is the circumburst environment and the interstellar medium of the host galaxy.  Although it is not possible to make a precise estimate of this component, the burst rotation measure, the amount of Faraday rotation exhibited by linearly polarized emission caused by its propagation through a magnetized plasma, presents a means of identifying those bursts whose radiation has likely propagated through a substantial ($>100\,\mdmunits$) amount of matter in the host galaxy.  For each burst the Milky Way contribution to RM
for $|b| > 10\,$deg is small ($<250\,$rad\,m$^{-2}$) and measurable\citemain{oppermann+12} and the intergalactic medium contribution is estimated\citemain{akahori10} to be $\sim 1\,$rad\,m$^{-2}$.  
Galactic halos, similarly, have been inferred to make contributions of several tens of $\rmunits$ to the RM\citemain{bernet08} from radio-loud quasar observations,
but our first analysis with an FRB\citemain{Prochaskaetal19} 
yields RM~$< 10 \, \rmunits$.
A suitable cutoff due to host galaxy ISM contamination is suggested by assuming the host galaxy magnetic field strength is comparable to that of our Galaxy. Measurements of Faraday rotation and dispersion from pulsars in the Milky Way (viz. figure 3 in\citemain{Ravi16}) exhibit a mean trend ${\rm DM}  = 1.55 |{\rm RM}|^{0.95} \equiv f_{\rm DM}({\rm RM})$, where RM and DM are measured in their usual units of rad\,m$^{-2}$ and pc\,cm$^{-3}$ respectively.  We find that that root-mean-square deviation of the actual DM values from their values predicted on the basis of this trend using $|{\rm RM}|$ are 69\% of the DM (i.e. the rms errors are 69\% of the mean DM value: $\langle [{\rm DM} - f_{\rm DM}({\rm RM})]^2/{\rm DM}^2 \rangle^{1/2} =0.69$).  We further find that there is an 85\% probability that the actual DM deviates from its predicted value by less than 0.9 times the actual DM value, and a 96\% probability that the predicted value differs by less than 2.0 times the DM value. 
We therefore suggest that a cutoff criterion $|{\rm RM} - {\rm RM}_{\rm MW}|_{\rm observed} < 100\,(1+z)^2\,$rad\,m$^{-2}$ bounds the dispersion measure to ${\rm DM} \lesssim 250\,(1+z)\, \mdmunits$ with 85\% confidence. 

A similar trend observed between the DM and the temporal smearing of Galactic pulsars caused by scattering\citemain{Bhatetal2004,Krishnakumaretal2015} can also be used to place upper bounds on the host contribution. Recent updates to this relation\citemain{Krishnakumaretal2015} indicate that, on average, a pulse smearing time, $\tau$, less than 33\,ms (2\,ms) limits DM to $<200\,$pc\,cm$^{-3}$ ($300\,\mdmunits$) at 0.327\,GHz (1\,GHz). However the DM-$\tau$ relation exhibits $\approx 0.8\,$dex variation about the trend (as discussed in the context of FRBs in the supplementary material in \citemain{Bannister19}), thus requiring $\tau < 5\,$ms  to ensure a reasonable ($\approx 70$\%) confidence that the DM contribution is less than 200 \dmunits.   We caution that the use of $\tau$ as an indicator of the host galaxy DM contribution is subject to considerable uncertainty, since neither the distances to the scattering material from the bursts, nor even the nature of the turbulence responsible for the temporal smearing observed in FRBs is well established.  The estimates presented here would be invalid, for instance, if the scattering were associated with the direct burst environment rather than the interstellar medium of the host galaxy.

Adopting the above criteria to the current set of FRBs with redshift
estimates based on their association to galaxies
(Table~\tabFRBs), we eliminate the following sources
from cosmological analysis: \\
\noindent
{\it FRB\,121102}:  We exclude the repeating FRB\,121102 \citemain{Chatterjeeetal2017} from our analysis for two reasons: (a) the rotation measure of this burst is anomalously high\citemain{Michilli18}, being three orders of magnitude higher than other FRBs in this sample and indicating that this burst DM is likely contaminated by an abnormally high circumburst or host galaxy contribution, and (b) its location within 2 deg of the Galactic plane imparts a significantly larger and likely less well constrained DM contribution from the Milky Way relative to the high Galactic latitude bursts detected by ASKAP.  \\
\noindent
{\it FRB\,190523}: We have conducted the analysis both with and without FRB\,190523. The host galaxy identification\citemain{Ravi19} from the larger, $3" \times 8"$ localization region, is more uncertain than the ASKAP FRB detections and was partially based on an assumed DM-redshift relation which presents a potential source of bias in our analysis. \\
\noindent
{\it FRB\,171020}: The identification of the host galaxy associated with FRB\,171020 is predicated on a search volume confined to a specific distance based on an assumed DM-redshift relation\citemain{Mahony18}, and is therefore excluded from the sample.  Moreover, it is difficult to ascribe a numerical value to the likelihood of a correct association in this instance. \\
\noindent
{\it FRB\,190611}: Our follow-up observations for FRB~190611 identify
a galaxy at \hgC\ with redshift $z= \zfrbC$ offset by $\approx 2''$ from the current
estimate of the FRB localization.
The significant offset ($\approx 10$\,kpc at that redshift)
and large systematic uncertainty in the FRB localization and the presence of a closer, faint source revealed by deep GMOS i-band imaging preclude a secure association at this time.  
As with FRB~190523, we conduct our analysis both with and without this burst in our sample.

\subsection{Imaging and Astrometry of FRBs~\listFRBs }
\label{sec:astrom}

The procedure for characterising the position and positional uncertainty of FRBs \listFRBs\ followed that described in the Supplementary Material of\citemain{Bannister19,Prochaskaetal19}.  For the purposes of extracting these observables, we use only the total intensity data.

For each FRB, raw voltage data for a suitable calibrator source was captured via the CRAFT pipeline in the hours following the burst detection.  For FRB~190102 and FRB~190608, the source PKS~1934-638 was used, while for FRB~190611 and FRB~190711, it was PKS~0407-658.  From these calibrator data and the FRB data, visibility datasets were produced using the DiFX correlator\citemain{Deller11}.  An initial coarse search for the FRB position used the dispersion measure, pulse duration, and approximate position from the incoherently summed FRB detection data, and after detection in the interferometric data a re-correlation was performed with revised position, dispersion measure, and pulse time/duration.  Radio Frequency Interference (RFI) was mitigated for the FRB dataset by subtracting visibilities from an adjacent time range surrounding the burst itself.  Additionally, for each FRB a visibility dataset and image was generated using the entire 3.1s of raw voltage data, to identify background radio continuum sources whose positions could be compared to catalogue values and verify the astrometric accuracy.

Per-station frequency-dependent complex gain calibration was derived from the calibrator dataset using the ParselTongue\citemain{Kettenis06} based pipeline described in \citemain{Bannister19} and transferred to the FRB datasets, prior to imaging in the Common Astronomy Software Applications (CASA) package. Best-fit positions and uncertainties were the extracted for each source using the task JMFIT in the Astronomical Image Processing System (AIPS)\citemain{Greisen03}.

Statistical uncertainties on the FRB positions were less than 0.5'' in all cases.  However, as discussed in \citemain{Bannister19} and\citemain{Prochaskaetal19}, the phase referenced FRB images will be subject to a systematic positional shift resulting from the spatial and temporal extrapolation of calibration solutions.  The magnitude of this systematic shift can  be estimated by comparing the positions of continuum sources in the field surrounding the FRBs to their catalogue values.  The accuracy to which this can be performed depends on the number of continuum sources visible in the ASKAP continuum image and their brightness, as well as the degree to which their intrinsic source structure can be modelled (or neglected).  For any given continuum source, the presence of unmodelled structure will act to shift the position of the source centroid and results in a measured offset between the ASKAP and reference positions, which perturbs the actual systematic positional shift.  However, the direction of such a shift depends on the source structure, and hence should not be correlated between different continuum sources.  For FRB~190102, FRB~190611, and FRB190711, observations made with the Australia Telescope Compact Array at a comparable frequency and angular resolution to the ASKAP image minimise the impact of source structure, but for FRB~190608, we made use of the Faint Images of the Radio Sky at Twenty centimetres (FIRST) survey\citemain{Becker95}, which has angular resolution roughly twice that of the ASKAP images.

Assuming the phase referencing errors result in a simple translation of the FRB field image, we estimate the magnitude of this offset and its uncertainty with a weighted mean of the measured offsets for each of the continuum sources in the FRB field, after discarding any sources that were resolved in either the ASKAP image or the reference image.  The magnitude of the offset ranged between 0 and 1.7 arcseconds, with uncertainties ranging from 0.3 to 0.6 arcseconds.


\noindent
\subsection{Optical identification and spectroscopy of the FRB host galaxies}
\label{sec:OptSpec}

The optical spectroscopy and redshift determinations for FRB\,180924 and FRB\,181112 have been outlined previously\citemain{Bannister19,Prochaskaetal19}. Spectroscopy of the host galaxies of FRB\,190102 and FRB\,190611 was conducted using the FOcal Reducer and low dispersion Spectrograph 2\citemain{FORS2} (FORS2) on the European Southern Observatory's Very Large Telescope (VLT) on Cerro Paranal. FORS2 was configured with the GRIS\_300I grism, an OG590 blocking filter, and a $1.3''$ wide slit, yielding a resolution $R_{\rm FWHM} \approx 550$. For FRB\,190102 $2\times600$~sec exposures were obtained on 2019~March~25~UT, while for FRB\,190611 $2\times1350$~sec exposures were taken on 2019~July~12~UT. These and associated calibration images were processed with the PypeIt software package\citemain{PypeIt}
to derive flux and wavelength calibrated spectra.

For the host galaxy of FRB\,190608, the optical spectrum from 
the 7th data release (DR7) of the Sloan Digital Sky 
Survey\citemain{sdssdr7} (SDSS)
was retrieved from the {\sc igmspec} database\cite{igmspec}.

Imaging of the host galaxies of FRB\,180924, FRB\,181112 and FRB\,190102 was undertaken using FORS2 on the VLT, while the FRB\,190608 and FRB\,190711 hosts were imaged with VLT/X-shooter\citemain{Xshooter}.
Imaging of the host galaxies of FRB\,190611 and FRB\,190711 was undertaken using GMOS on Gemini-South\cite{GMOS-S}, from sets of 44 and 12 images of 100\,sec each in the $i$-band, respectively. 

The FORS2 images were first reduced with ESO Reflex\citemain{ESOReflex}, further processed in Python, and then co-added using a median combine in Montage\citemain{Montage}. The WCS solutions were updated with Astrometry.net\citemain{Astrometry}, with further adjustments performed by comparison with \textit{Gaia}\cite{Gaia} or Dark Energy Survey\citemain{DES} positions. The X-shooter images were reduced using a custom Python pipeline making use of the package {\sc ccdproc}\citemain{ccdproc}, including measures to cope with prominent fringe patterns in $I$-band; the images were then co-added and the astrometry adjusted with the same method as above. The GMOS images were reduced and co-added with {\sc Pyraf} using standard procedures; the astrometry was adjusted with the same method stated above.
Projected distances were estimated using Ned Wright's Javascript Cosmology Calculator\citemain{Wright2006}.

Two of the FRBs in the gold-standard sample, FRB 190608 and 190711, have offsets larger than 1\,arcsec from the galaxy light centroid.  FRB 190608, however, is a $z=0.11$ galaxy (i.e.\,\,nearby) and the chance projection is even less than 0.3\%.  Regarding FRB 190711 we estimate a $6.1 \times 10^{-3}$ probability that an unrelated galaxy is within the error circle+FRB-galaxy centroid distance (for the measured $R(AB)=23.7 \pm 0.2$\,mag as calibrated against the SkyMapper survey), and we estimate a probability $p=1.9 \times 10^{-3}$ for an unrelated galaxy to be within the FRB error circle but below the detection limit of $r=25.5$\,mag. The remainder of the host galaxy associations for each FRB have a probability $P < 10^{-3}$ of a chance occurrence\citemain{Prochaskaetal19}.

The radio burst dynamic spectra and host galaxy optical spectra are shown in \figfrbA.


\subsection{Estimating \dmacosmic}
\label{sec:DM_cosmic}

Central to the analysis is an estimate of the average
\dmcosmic\ value as a function of redshift and for
a given cosmology, as defined in Equation~\ref{DMzRelation}.
Previous formulations\citemain{Inoue04,deng2014} have adopted
similar definitions but with less precise considerations
for $f_d(z)$, the fraction of cosmic baryons in diffuse
ionized gas.  Our formulation considers the redshift evolution
of three dense baryonic components that will not
contribute to $n_e$:
(1) stars,
(2) stellar remnants (e.g.\ white dwarfs, neutron stars),
(3) the neutral ISM of galaxies.
For (1), we interpolate the empirically estimated
stellar mass density estimates\citemain{md14}.
For (2), we adopt the estimation of \citemain{fukugitaandpeebles04}
which is 30\%\ of the stellar mass.
For (3), we assume the mass ratio of the ISM to stars
is constant from $z=0-1$ and adopt the present-day
estimate\citemain{fukugitaandpeebles04} of
$M_{\rm ISM}/M_* = 0.38$.
The model also allows for the partial ionization of helium
but this is not relevant for the FRBs considered here.
All of these calculations are encoded in Python
in the public FRB repository
(https://github.com/FRBs/FRB).  Censuses of the gas and star evolution of baryons in $z<1$ systems constrain the error in the fraction of neutral and non-diffuse baryons (i.e. $1-f_d(z)$) to $\approx 30\%$ at present.  Thus, with this component constituting $\sim 15\%$ of the total baryon budget at $z\approx 0$,  the correction to \obh is uncertain at a level below 6\%, well below the level of precision that investigation of the current FRB sample permits.  We refer the reader to \citemain{2019arXiv190902821W} for a discussion on the constraints on $f_d(z)$ possible in future with a larger sample of FRBs.

\subsection{Cosmological and host galaxy parameter estimation}
\label{sec:cosmology} 

The ASKAP FRB measurements and localizations
afford a new opportunity to constrain our
cosmological paradigm through estimations of 
\dmcosmic\ and \zzfrb.
The cosmic DM is governed primarily by the baryonic density $\Omega_b$ and the expansion rate of the Universe, $H_0$, and the fraction of baryons in the diffuse phase, $f_d(z)$.  In the following, we will
assume a flat cosmology with $\Omega_\Lambda = 0.691$
(Planck15). The expansion rate is dominated by this dark
energy term for $z < 0.7$, so cosmological analysis of the ASKAP
FRBs is not sensitive to the precise value of 
$\Omega_m$ and, therefore, to a close approximation,
$\mdmacosmic \, \propto \, \Omega_b H_0$. 
We therefore proceed to place a constraint on this
product.

To construct a likelihood function \like\ from our FRB
measurements, we build a model for \dmcosmic\ and
its uncertainty.  The model
is based primarily on the cosmological parameters,
but it must also 
allow for a nuisance parameter which accounts for
the DM of our Galactic halo and that of the host galaxy:
$\mdmmw + \mdmhost$.
For the former term, theoretical models informed by observation
suggest $\mdmmw \approx 50 \mdmunits$ with a small
dispersion\citemain{deng2014} \citemain{Prochaska19}, but
we acknowledge that the mean value is poorly constrained.
We expect the variance in these terms to be
driven by \dmhost, which follows from the large range
in DM values observed for the ISM of our Galaxy \citemain{NE2001},
even if one ignores whether FRBs occur in `special'
locations within a galaxy.
Furthermore, the very high RM and (likely related) large 
DM excess of FRB~121102 above \dmcosmic\ implies at least
one FRB with a large \dmnui\ value \citemain{Michilli18}.

The PDF for \dmnui\ has limited theoretical motivation.  In the following, we
assume a log-normal distribution which has two salient features:
(1) it is positive definite;
(2) it exhibits an asymmetric tail to large values.
The latter property allows for high \dmnui\ values that might arise from
gas local to the FRB, e.g.\ an HII region or 
circumstellar medium.  Formally, we adopt a log-normal distribution  
\begin{equation}
\mphost = \frac{1}{(2 \pi)^{1/2} {\rm DM} \sigma_{\rm host}} \exp \left[ - \frac{(\log{\rm DM}-\mu)^2}{2 \sigma_{\rm host}^2} \right] 
\label{eqn:dmhost}
\end{equation}
This distribution has a median value of $e^{\mu}$ and variance $e^{\mu +\sigma_{\rm host}^2/2} ( e^{\sigma_{\rm host}^2} - 1 )^{1/2}$.  We consider distributions with $e^{\mu}$ in the range $20-200\, \mdmunits$ and $\sigma_{\rm host}$ in the range $0.2-2.0$.  An illustrative set of these probability distribution functions 
for \dmnui\ is shown in \fighostDM.  For consistency of interpretation of ${\rm DM}_{\rm host}$ values from bursts at disparate redshifts, the probability distribution function is referenced to the rest frame of the host galaxy, so a correction ${\rm DM}_{\rm host} \rightarrow {\rm DM}_{\rm host} \, (1+z_{\rm FRB})^{-1}$ is applied and the distribution normalised accordingly, however, in practice this redshift correction factor varies only over the range 0.7 to 0.9 in the gold-standard sample.  The inferred dispersion in ${\rm DM}_{\rm host}$ is consistent with the expected range of host DMs given the galaxy type, morphology and orientation on the sky and the distance of the FRB from the galaxy centre.  However, we are unable to state more than this at present, and remark that present estimates of the ${\rm DM}_{\rm host}$ contributions towards specific localised FRBs\citemain{Bannister19,Marcote2020} with two quite different host galaxies are, respectively, in the range $30$-$81\,\mdmunits$, and $<70\,\mdmunits$. This suggests that any correction on this basis could be small for our sample.  A further interesting aspect of our measurements is that it is beginning to place limits on these corrections. 

Altogether,
$\mdmfrb^{\rm model} = \mdmcosmic(z) + 
\mdmnui + \mdmism$ with the latter
quantity estimated from NE2001 based on the FRB coordinates; given the high Galactic latitudes of the present ASKAP sample we adopt a value $\mdmism=30 \mdmunits$ for these bursts.  The mechanics of our treatment of the ${\rm DM}_{\rm host}$, ${\rm DM}_{\rm MW,halo}$ and ${\rm DM}_{\rm MW,ISM}$ terms is described in greater detail in eq.(6).

The model probability distribution for \dmcosmic\ is derived from theoretical treatments of the IGM and galaxy halos \citemain{McQuinn14,Prochaska19} with \sdm\ dominated
by the physical variance in \dmcosmic.  \figMoreModels\ shows that comparison against the analytic form (as used in other IGM-related contexts \citemain{MHR00}), 
\begin{eqnarray}
p_{\rm cosmic}(\Delta) = A \Delta^{-\beta} \exp\left[ - \frac{(\Delta^{-\alpha} - C_0)^2}{2 \alpha^2 \msdm^2} \right], \qquad \Delta > 0,
\label{eqn:MHR}
\end{eqnarray}
provides an excellent match to the \dmcosmic\ distributions observed in our semi-analytic models and in a hydrodynamic simulation, where $\Delta \equiv {\rm DM}_{\rm cosmic}/\langle {\rm DM}_{\rm cosmic} \rangle$.  The motivation for this form is that in the limit of small \sdm, the distribution of ${\rm DM}$ should approach a Gaussian owing to the Gaussianity of structure on large scales (a significant component of the variance of $p_{\rm cosmic}(\Delta)$ comes from tens of megaparsec structures) and in the low-$\sigma_{\rm DM}$ limit the halo gas is more diffuse and so the PDF approaches a Gaussian owing to the intersection of the line of sight with more structures.  Conversely, when the variance is large, this PDF captures the large skew that results from a few large structures that contribute significantly to the ${\rm DM}$ of many sightlines. The sharp low-DM cutoff in the distribution reflects the fact that a significant component of the IGM is highly diffuse, and displays much less variance than the halo-related component, thus imposing a strict lower limit to the DM.  The parameter $\beta$ is related to the inner density profile of gas in halos. If the 3D density profile scales as $\rho\propto r^{-\alpha}$, $\beta = (\alpha+1)/(\alpha-1)$ such that an isothermal profile with $\alpha=2$ has $\beta=3$ and an inner slope of $\alpha=1.5$ has $\beta=5$.  Such slopes are consistent with those found in numerical simulations of intrahalo gas \citemain{2019MNRAS.488.1248H,2017MNRAS.466.3810F}.  The indices $\alpha=3$ and $\beta=3$ provide the best match to our models (although we find that $p_{\rm cosmic}(\Delta)$ is weakly sensitive to order unity changes in these parameters, with $\beta=3$ having the most flexibility for our $z=0.11$ measurement relative to $\beta=4$). 
We use the parameter \sdm\ in $p_{\rm cosmic}(\Delta)$ as an effective standard deviation even though formally the standard deviation with $\beta=3$ diverges logarithmically.  We find that \sdm\ is closely tied to the true standard deviation when imposing motivated maximum cutoffs for $\Delta$ on the distribution.  The mean of the distribution requires that $\langle \Delta \rangle = 1$, which fixes the remaining parameter $C_0$ in $p_{\rm cosmic}(\Delta)$.  

\figMoreModels\  shows models that use eq.~(\ref{eqn:MHR}) for $p_{\rm cosmic}$ relative to numerical calculations at redshifts that span the considered range. The solid curves are the previously described semi-analytic models\citemain{McQuinn14}, which assume that halos below the specified mass have been evacuated of gas, and the `swinds' simulation of \citemain{Faucher-Getal2011}.  The dashed curves show the function evaluated for the best-fit $\sigma_{\rm DM}$, and the dot-dashed curves adopt the parameterization $\msdm = F z^{-0.5}$ and scale off the $z =0.5$ best fit value for \sdm\, yielding $F$ of $0.09$, $0.15$ and $0.32$ in our semi-analytic models in which halos of $10^{14}$, $10^{13}$ and $10^{11}M_\odot$ are evacuated of their gas.  The agreement of the dot-dashed curves with the solid numerical model curves demonstrates that $\msdm = F z^{-0.5}$ approximates the evolution over the range of our measurements. This scaling is further motivated in the Euclidean limit, applicable for $z \ll 1$, where $\langle {\rm DM}_{\rm cosmic} \rangle  = n_e c z/H$  and \sdm $\approx  {\rm DM}_{\rm halo} \sqrt{N} / \mdmacosmic$, where $n_e$ is the mean electron density and  $N$ is the number of halos intersected, which is proportional to the path length probed or $c z/H$. 

While our analytic parameterisation describes the distribution of \dmcosmic\ both in semi-analytic models and numerical simulations, we use the more flexible semi-analytic models to set the marginalization range in $F$ that is used for some constraints on \obh.  Here we argue that the considered semi-analytic models shown in \figMoreModels\ span the likely range of possible feedback scenarios.  These models approximate halos as retaining their gas in a manner that traces the dark matter above some mass threshold.  This approximates the picture in many simulations\citemain{fielding+17, 2019MNRAS.488.1248H} and analytic models\cite{Sharma+12, Voit+19} in which the fraction of halo gas retained is a strongly increasing function of halo mass prior to saturating at unity.  Furthermore, gas that is outside of halos is less effective at contributing variance: take the example where gas is distributed out to a distance $R$ around a halo.  The probability a sightline intersects this gas scales as $R^2$, leading to less shot noise for larger $R$, while the contribution of each individual system scales as $R^{-2}$, leading to a smaller contribution for larger $R$.  This picture motivates the semi-analytic model's approximation that ejected gas diffusely traces large-scale structure\citemain{McQuinn14}.  

Simulations and models generally find that halos below thresholds masses in the range $\sim 10^{11}-10^{13}M_\odot$ are evacuated of gas\citemain{Sharma+12, fielding+17, 2019MNRAS.488.1248H, Voit+19}, although some implementations of stellar quasar feedback can result in significantly different predictions\citemain{2019MNRAS.484.1637J}.  Halo gas in $M >10^{14}M_\odot$ halos is constrained by X-ray observations to mostly reside within such halos\citemain{2019arXiv190805765D}.  Our strongest feedback model, in which $F=0.09$, pushes up against this observational limit. Our model with the weakest feedback assumes that dwarf galaxy-sized halos with $10^{11}M_\odot$ retain their gas and yields $F=0.32$ (and we find that $F$ is just marginally larger if the $10^{10}M_\odot$ halos of the smallest dwarf galaxies retain their gas, halos just massive enough to overcome the pressure of the intergalactic medium and retain their gas\citemain{2014MNRAS.444..503N}).  Thus our models span the range of likely feedback scenarios. 




Given this semi-analytic formalism, we proceed to estimate the model likelihood by computing the joint likelihoods of all FRBs:
\begin{equation}
 \mlike = \prod\limits_{i=1}^{N_{\rm FRBs}} P_i (\mdmpfrb|z_i), 
\label{eqn:L}
\end{equation}
where  $P_i (\mdmpfrb|z_i)$ is the probability 
of the total observed \dmfrb\ corrected for the Galaxy:
\begin{equation}
\mdmpfrb \equiv \mdmfrb - \mdmism -
\mdmmw = \mdmnui + \mdmcosmic 
\label{eqn:dmpfrb}
\end{equation}
For a burst at a given $z_i$ and the model parameters we have:
\begin{equation}
P_i({\rm DM_{\rm FRB, i}'| z_i})  =  \int\limits_0^{\rm DM_{\rm FRB}'} \;
  \mphost \; p_{\rm cosmic}({\rm DM_{\rm FRB, i}' - \mdmnui}, z_i) \; d\mdmnui  \;\; ,
\label{eqn:prob}
\end{equation}
and
\phost\ the PDF for \dmnui.
With the likelihood function defined we construct a grid
of $\Omega_b H_0$, $F$, $\mu$ and $\sigma_{\rm host}$ values and marginalize over
the last three to obtain the constraint on \obh.  These results are presented in Figure~\figobh\ in the main text for the gold-standard sample. 
\figsevenburstcorner\ presents the results of the same analysis when FRBs 190523 and 190611 are included in the dataset.

To place confidence intervals on $\Omega_b H_0$ and \sdm, we use the likelihood ratio test statistic $\mathcal{D}$:
\begin{eqnarray}
\mathcal{D} (\Omega_b h_{70}, F, \mu, \sigma_{\rm host}) = 2 \log {\mathcal L}_{\rm max} - 2 \log {\mathcal L}(\Omega_b h_{70}, F, \mu, \sigma_{\rm host}),
\end{eqnarray}
where $\mathcal{L}_{\rm max}$ is the maximum value of $\mathcal{L}$, i.e.\ for parameters maximising the likelihood. According to Wilks' theorem, for a sufficiently large number of FRBs, $\mathcal{D}$ will be distributed according to a $\chi^2_n$ distribution with $n=4$ degrees of freedom\citemain{wilks1938}. If the cumulative distribution function of the $\chi^2_4(x)$ distribution is CDF$(x)$,  solving CDF$(x)=p$ constrains the $\Omega_b h_{70}, F, \mu, \sigma_{\rm host}$ parameter space to the region $\mathcal{D} \le x$ at confidence level (C.L.) $p$.

Uncertainties in these confidence estimates are likely dominated by systematic effects in the sample selection, and small number statistics. To test both, we extend the gold-standard sample of five bursts to include FRBs 190523 and 190611. The resulting analysis is shown in \figsevenburstcorner. Compared with Figure~\figobh, the inclusion of two further bursts shifts the maximum-likelihood estimate for $\Omega_b h_{70}$ at 68\% C.L.\ from $0.051_{-0.015}^{+0.014}$ to $0.042_{-0.012}^{+0.011}$, i.e.\ consistent with the original uncertainties. This does not mean that there is no systematic bias, nor that Wilks' theorem holds precisely for our sample, but rather that any such effects are minor compared to the inherent uncertainties from our small sample size of localized bursts.

\subsection{Accounting for biases in the probability distribution}
\label{sec:supp_biases}

The cosmological evolution of the FRB population, and its intrinsic luminosity function, can strongly influence the observed/expected distribution of FRBs in redshift--DM space\citemain{Macquart18c}, P$({\rm DM},z)$. We therefore perform our likelihood maximisation over  P$({\rm DM}| z)$ only. This discards the information contained in the redshifts of our detected FRBs, but makes the procedure more robust against factors influencing the redshift distribution.

The remaining bias comes from changing sensitivity as a function of DM. This can be either direct, through DM-smearing within frequency channels, or indirect, through increased scatter broadening associated with the same gaseous structures causing the observed DM.

We wish to compute the dispersion measure limit, ${\rm DM}_{\rm cutoff}$ at which a given FRB would have been undetectable.
The S/N of a detected burst depends on its intrinsic (or scatter-broadened) width, $w$, the time resolution of the detection system, $t_{\rm res}$, and the amount of dispersion measure time smearing between adjacent 1\,MHz spectral channels, $t_{\rm smear}({\rm DM})$.  The resulting width of the pulse is 
\begin{eqnarray}
\Delta t_{\rm obs}({\rm DM}) = \sqrt{w^2 + t_{\rm res}^2 + t_{\rm smear}^2({\rm DM})}.
\end{eqnarray}
We compute ${\rm DM}_{\rm cutoff}$ such that the burst, detected by our system with a signal-to-noise ratio of $s_0$ at a ${\rm DM}_{\rm obs}$ would have fallen below our detection threshold of $s_{\rm d} = 9.0 \sigma$.  For each burst we thus solve
\begin{eqnarray}
s_d = s_0 \sqrt{\frac{\Delta t_{\rm obs}({\rm DM}_{\rm obs})}{\Delta t_{\rm obs}({\rm DM}_{\rm cutoff})}}
\end{eqnarray}
\tabDMcutoff\ lists the DM, widths, time resolution, detection S/N values and derived ${\rm DM}_{\rm cutoff}$ values for each of the bursts in our sample.

\subsection{MCMC Analysis}
\label{sec:mcmc}

To complement the likelihood analysis presented in the 
main text, we have performed Bayesian inference of a model
constructed to describe the DM and redshift measurements of the FRBs.
The model consists of four parameters describing two probability distribution
functions for distinct components of the dispersion measure:
 (i) \dmcosmic, which describes the extragalactic dispersion measure including
 both the diffuse IGM and the gas associated with intervening galactic halos; and
 (ii) \dmnui, which describes ionized gas associated with the host galaxy (we assume
 a fixed \dmmw\ value of 50\,\dmunits\ for the Galactic halo).
We parameterize the former PDF with Equation~\ref{eqn:MHR}, i.e.
$p_{\rm cosmic}(\Delta)$
with $\Delta \equiv \mdmcosmic / \mdmacosmic$ and 
\dmacosmic\ the average value for the assumed cosmology 
(Equation~\ref{DMzRelation}).
The foregoing subsection on cosmology and host galaxy parameter estimation describes theoretical treatments that motivate one to
adopt $\alpha=3$ and  $\beta=3$ in Equation~\ref{eqn:MHR}
and to adopt the functional form of
$\sigma_{\rm DM} = F / z^{1/2}$ for its 
dispersion parameter.
For \dmacosmic, we modulate its 
amplitude via the product \obh.
Therefore, \pcosmic\ is governed by two free 
parameters: $F$ and \obh.

We adopt the same \phost\ PDF described earlier,
%
with free parameters $\exp(\mu)$ and $\mshost$.
%
%
From these two PDFs we construct a likelihood function for the set
of observed FRBs using Equations~\ref{eqn:L}
and \ref{eqn:prob}.
Note that measurement uncertainty in \dmfrb\ does not
enter into the evaluation of $\mathcal{L}$ because the
dispersion from \dmcosmic\ and \dmnui\ are much greater.
Put another way, our model is constructed
to describe the observed distribution of \dmpfrb\ values
with an anticipated dispersion substantially exceeding
the uncertainty in individual \dmfrb\ measurements
(typically $<1\mdmunits$).

Effectively, two of the parameters (\obh, $\mu$) set the amplitude
of the DM-$z$ relation and two describe its dispersion
($F, \sigma_{\rm host}$).  We anticipate a degeneracy between each set
although if \dmhost\ is approximately independent of redshift
then this apparent degeneracy may be resolved.
Only the dispersion in \dmcosmic, parameterized
by $F$, allows for large negative excursions from the mean relation.
Lastly, we introduce priors for the four parameters based on
a combination of experimentation, physical expectation, and
scientific motivation.
For \obh, the scientific focus of this manuscript,
we adopt a uniform prior ranging from $0.015-0.095$ 
which easily spans the Planck15 estimate.
For $F$, we adopt a uniform prior in the interval $(0.01, 0.5)$, a larger range than anticipated by our models in the frequentist analysis presented above. Regarding $\exp[\mu]$, we adopt a uniform prior in the
interval $[20, 200] \, \mdmunits$.  We consider lower values for
the mean to be non-physical and we will find that larger values
are disfavored by the observations.
Lastly, we assume a uniform prior for $\sigma_{\rm host}$ in
the interval $[0.2, 2]$.  The larger $\sigma_{\rm host}$
values give non-negligible probability for \dmhost\ values in
excess of 1000\,\dmunits.
Future observations, especially an ensemble of FRBs at low redshift,
will better inform these priors on $\mu$ and $\sigma_{\rm host}$.

Adopting the above likelihood and priors, we performed a Bayesian
inference of the four parameters using the gold sample of FRB
measurements and standard MCMC techniques.
These were performed with the {\sc pymc3} software package using
slice sampling and four independent chains of 40,000 samples after
a tuning period of 2,000 samples.  
\figmcmccorner\ presents a corner plot of the combined samples.
A principal result is that the data yield a \obh\ distribution
fully consistent with the independent estimates from the CMB,
BBN, and supernovae.  
Quantitatively, the \obh\ samples have a median value of
\vobh\ and a 68\%\ confidence interval spanning
\iobh\  (see \tabMCMC).
Taken strictly, at 95\%\ confidence these FRB measurements require
a universe with at least 70\%\ of the baryons inferred from 
BBN and CMB analysis.
These results hold despite the weak priors placed on the PDF for \dmhost,
but we caution that they are dependent on the value assumed
for \dmmw.

\figmcmccorner\ also reveals the anticipated anti-correlations
between $\mu$ and \obh\ and (to a lesser extent) $F$ and $\sigma_{\rm host}$.
We expect these to weaken as the FRB sample grows in size and redshift 
range.  Lastly, we note that the $F$ and $\mu$ parameters have maximal
probability at one edge of their assumed prior intervals.
 Values of $F$ that are on the higher side of the considered range (a range that spans the possible model space) are modestly favored.  For $\mu$, we consider 20\,\dmunits\ to be the lowest sensible mean
contribution from the host galaxy (which could also mean a lower value for the Galactic halo than adopted here).

The frequentist analysis in the main text and this Bayesian MCMC analysis agree very well on the gold sample.  The most notable differences are that the MCMC analysis prefers a distribution for $\exp[\mu]$ that is more peaked to smaller $\exp[\mu]$ values and one for $F$ that peaks towards larger values, although with no value for $\exp[\mu]$ or $F$ strongly preferred by either analysis.  When the parameters are not well constrained one would not expect perfect agreement between the methods, as, for example, the Bayesian analysis is sensitive to our prior on $\exp[\mu]$ when this parameter is not well constrained.  It is expected that the differences between the two methods will become smaller with more data.  Already for the seven burst sample (\figsevenburstcorner), the distribution for $F$ in the frequentist analysis is more similar to the MCMC analysis of the gold sample.

\end{methods}

\bibliographystylemethods{naturemag}
\setcounter{enumiv}{30}

\begin{addendum}

 \item  We thank H. Yang and L. Infante for providing IMACS imaging around FRB\,190611 that informed the further follow-up observations on VLT and Gemini-S presented here.
 We are grateful to Australia Telescope National Facility (ATNF) operations and the Murchison Radio-astronomy observatory staff for supporting our ASKAP operations, and the ATNF steering committee for allocating time for these observations. 
K.W.B., J.P.M, and R.M.S. acknowledge Australian Research Council (ARC) grant DP180100857
 A.T.D. is the recipient of an ARC Future Fellowship (FT150100415).
 S.O. and R.M.S. acknowledge support through ARC grant FL150100148. 
 R.M.S. also acknowledges support through ARC grant CE170100004. N.T. acknowledges
support from FONDECYT grant number 11191217 and PUCV/VRIEA project 039.395/2019.
The FRB detection pipeline makes use of the DBSCAN algorithm \citemain{ester_density-based_1996}, as implemented by\citemain{Novikov2019}, to mitigate RFI and reduce the frequency of false-positive FRB triggers.  The Australian Square Kilometre Array Pathfinder and Australia Telescope Compact Array are part of the Australia Telescope National Facility which is managed by CSIRO. 
Operation of ASKAP is funded by the Australian Government with support from the National Collaborative Research Infrastructure Strategy. ASKAP uses the resources of the Pawsey Supercomputing Centre. Establishment of ASKAP, the Murchison Radio-astronomy Observatory and the Pawsey Supercomputing Centre are initiatives of the Austrperformedalian Government, with support from the Government of Western Australia and the Science and Industry Endowment Fund. Part of this work was performed on the OzSTAR national facility at Swinburne University of Technology. OzSTAR is funded by Swinburne University of Technology and the National Collaborative Research Infrastructure Strategy (NCRIS).
We acknowledge the Wajarri Yamatji as the traditional owners of the Murchison Radio-astronomy Observatory site. 
This work includes observations collected at the European Southern Observatory under ESO programmes 0102.A-0450(A), 0103.A-0101(A) and 0103.A-1010(B).
This work includes data obtained from program GS-2019B-Q-132
at the Gemini Observatory, acquired through the Gemini
Observatory Archive and processed using the Gemini {\sc Pyraf} package.
Gemini Observatory is operated by the Association of Universities
for Research in Astronomy, Inc., under a cooperative agreement
with the NSF on behalf of the Gemini partnership: the National
Science Foundation (United States), National Research Council
(Canada), CONICYT (Chile), Ministerio de Ciencia, Tecnolog\'{i}a
e Innovaci\'{o}n Productiva (Argentina), Minist\'{e}rio da
Ci\^{e}ncia, Tecnologia e Inova\c{c}\~{a}o (Brazil), and Korea
Astronomy and Space Science Institute (Republic of Korea). {\sc Pyraf} is a product of the Space Telescope Science Institute, which is operated by AURA for NASA.

\item[Author Contributions]  J.X.P., M.M. and J.P.M. framed the analysis approach and drafted the manuscript, with contributions from A.T.D., R.D.E. and R.M.S.  K.W.B developed the detection and localization pipelines, and with A.T.D. and C.P. jointly developed the correlation code and interferometry processing pipeline.  D.R.S. made additional improvements to the performancy of the detection pipeline.  R.M.S. and S.B. detected the FRBs and performed follow-up astrometry.  A.T.D. and C.K.D. derived the FRB positions from the CRAFT voltage data, and S.R., J.X.P., N.T. and L.M. obtained and reduced the optical data to derive the burst host galaxy identifications and redshifts.  M.M., J.X.B. and J.P.M.  developed the IGM model and analysis code, and S.O. adapted the code to run on the Swinburne supercomputer and collated the results.  C.W.J. contributed to the maximum likelihood analysis and framed the approach to estimate parameter uncertainties.

\item[Author information]
\item[Competing Interests] The authors declare that they have no competing financial interests.


\item[Correspondence] Correspondence and requests for materials should be addressed to J.P.M.~(email: J.Macquart@curtin.edu.au) or J.X.Prochaska.~(email: xavier@ucolick.org)

 \item[Reprints] Reprints and permissions information is available at www.nature.com/reprints.

\end{addendum}

\subsection{Data availability}
The datasets generated during and/or analysed during this study are available at   \url{https://data-portal.hpc.swin.edu.au/dataset/observations-of-four-localised-fast-radio-bursts-and-their-host-galaxies}

\subsection{Code availability}
Custom code is available at \url{https://github.com/FRBs/FRB}

\clearpage

{\bf Extended Data}


\setcounter{figure}{0}   
\setcounter{table}{0}
\renewcommand{\figurename}{Extended Data Figure}
\renewcommand{\tablename}{Extended Data Table}

\begin{table}[htbp!]
\begin{longtable}[c]{llllll}
\hline
\hline
  \multicolumn{1}{c}{FRB} &  \multicolumn{1}{c}{detection $S/N^{(1)}$ } & \multicolumn{1}{c}{DM} & \multicolumn{1}{c}{$w$}  & \multicolumn{1}{c}{$t_{\rm res}$} & \multicolumn{1}{c}{${\rm DM}_{\rm cutoff}$} \\
  \multicolumn{1}{c}{} & \multicolumn{1}{c}{} & \multicolumn{1}{c}{(pc\,cm$^{-3}$)} &  \multicolumn{1}{c}{ms} &  \multicolumn{1}{c}{ms} & \multicolumn{1}{c}{(pc\,cm$^{-3}$)} \\
\hline
180924 & 21.1 & 361.42(6) & 1.76(9) & 0.864 & 3050 \\
181112 & 19.3 & 589.27(3) & 2.1(2) & 0.864 &  3400 \\
190102 & 14.0 & 363.6(3) & 1.7(1) & 0.864 &  1250  \\
190608 & 16.1 & 338.7(5) & 6.0(8) & 1.728 & 4510 \\
190611$^{(2)}$ & 9.3 & 321.4(2) & 2(1) & 1.728 &  430  \\
190711 & 23.8 & 593.1(4) & 6.5(5) & 1.728 & 11500 \\ 
\hline
\caption{\label{tab:DMcutoff} {\bf Detection properties of the ASKAP FRBs.}  The values of ${\rm DM}_{\rm cutoff}$ denote the maximum DM at which a burst with those properties listed would have been detectable at a S/N threshold, $s_t = 9.5$ with the ASKAP telescope backend at a centre frequency of 1295\,MHz given the burst width, $w$ and search time resolution, $t_{\rm res}$ and its 1\,MHz spectral resolution. 
(1) The detection $S/N$ value listed is that reported by the incoherent detection pipeline for the telescope beam in which the detection signal was strongest.
(2) The voltage-capture system enables the follow-up of sub-threshold events detected in the incoherent pipeline, and subsequent interferometric validation, which would increase the S/N of a valid event by a factor $\gtrsim 5$.  The reported ${\rm DM}_{\rm cutoff}$ is referenced to the threshold $s_t=9.0$ relevant to the observing run during which this event was detected.
}
\end{longtable}
\end{table}

\bigskip \bigskip

\begin{table}[htbp!]
\centering
\caption{{\bf Results of the MCMC Analysis.} The 68\%- and 95\%-confidence parameter range estimates from this analysis are consistent with the results of the approach described in the main text. \label{tab:mcmc}}
\begin{minipage}{270mm} 
\begin{tabular}{cccccc}
\hline 
Parameter & Unit & Prior & Median & 68\% & 95\%
\\ 
\hline 
$F$ & None & $\mathcal{U}$(0.011,0.5) & 0.31 &0.15,0.44 &0.04,0.49 \\ 
$\exp(\mu)$ & pc cm$^{-3}$ & $\mathcal{U}$(20.0,200) & 68.2 &33.2,127.8 &22.0,181.1 \\ 
$\sigma_{\rm host}$ & None & $\mathcal{U}$(0.2,2) & 0.88 &0.43,1.53 &0.24,1.91 \\ 
$\Omega_b h_{70}$ & None & $\mathcal{U}$(0.015,0.095) & 0.056 &0.046,0.066 &0.038,0.073 \\ 
\hline 
\end{tabular} 
\end{minipage} 
\end{table} 


\begin{figure}[htpb!]
\includegraphics[width=1.0\linewidth]{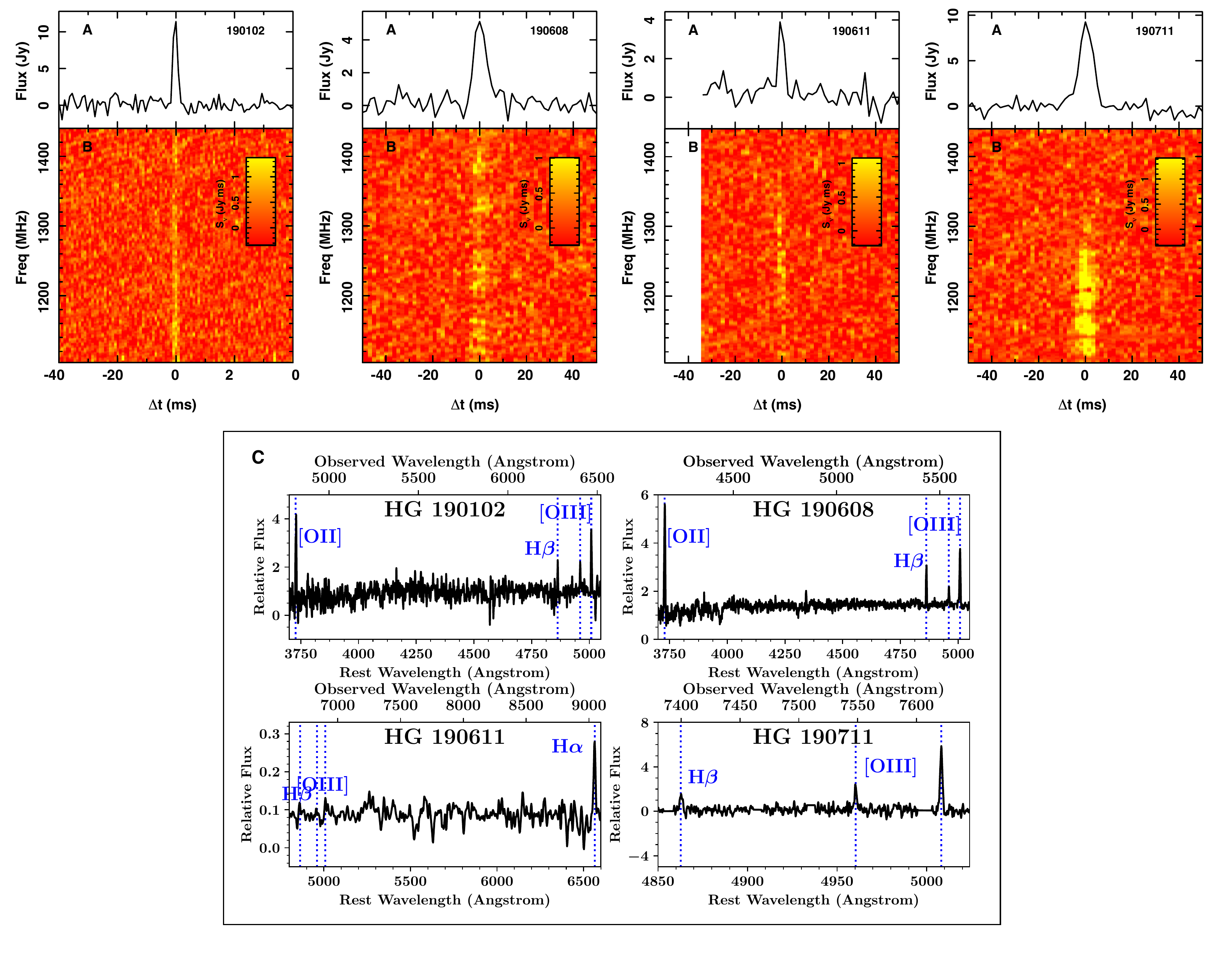}
\caption{{\bf The pulse profiles and host galaxy spectra of the four FRBs presented here.} The pulse profiles (panels A) and the radio dynamic spectra (panels B) showing the detections by the ASKAP incoherent capture system (ICS) of FRB\,190102 with a time resolution of 0.864\,ms, and FRBs\,190608, 190611 and 190711 with a resolution of 1.728\,ms. The spectral resolution is 1\,MHz across the 336\,MHz bandwidth.  Panels (C) show the SDSS (HG 190608) and VLT/FORS2 (HG 190102, HG 190611 and HG 190711) optical spectra of the host galaxies located at the respective FRB positions (see Table {\tabFRBs}), and the spectral lines from which their redshifts are deduced.} 
\end{figure}

\medskip

\begin{figure}[htbp!] \label{DMhostfig}
\includegraphics[width=\linewidth]{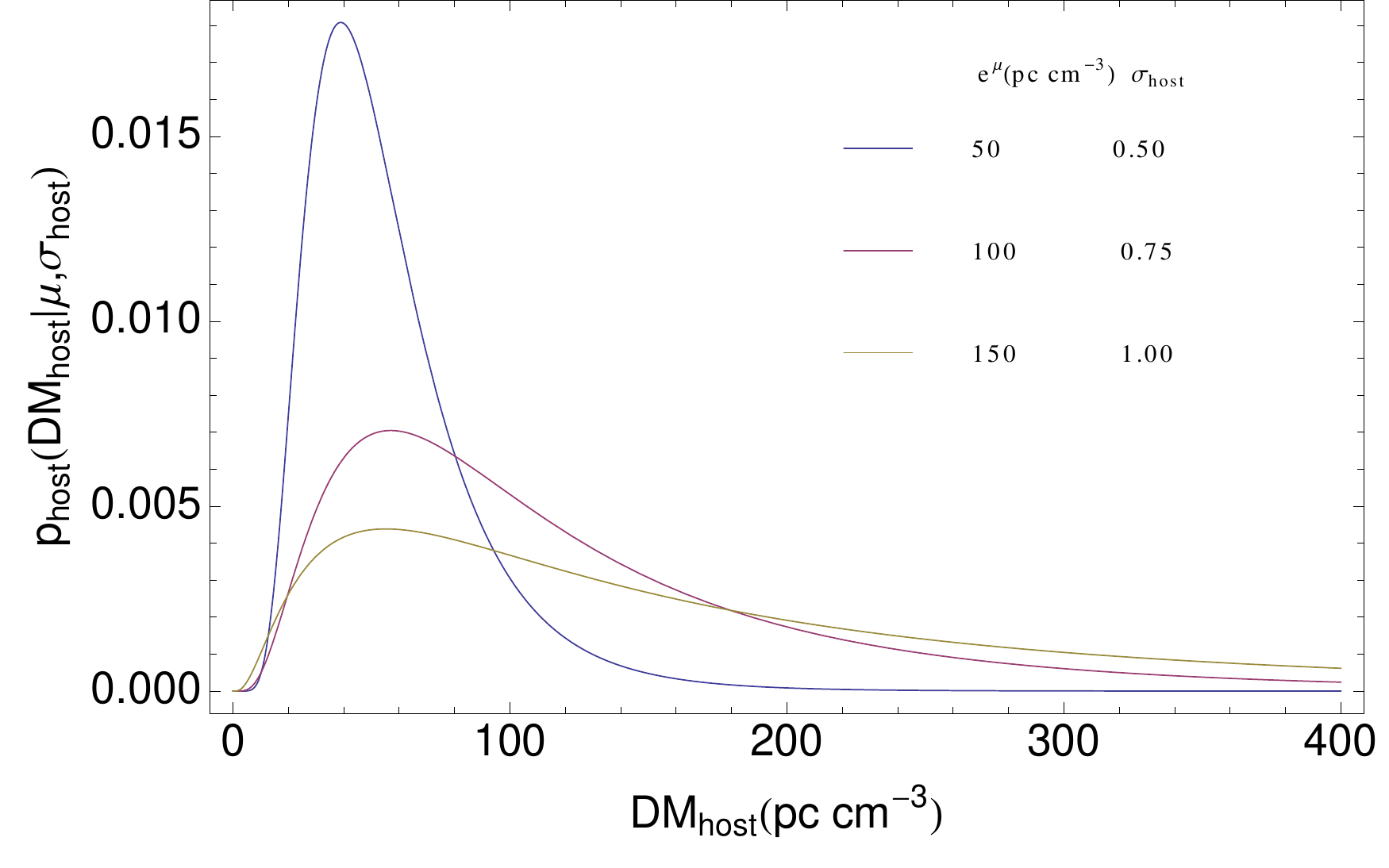}
\caption{{\bf The shape of the distribution used to model the host galaxy dispersion measure.} The behaviour of the probability distribution \phost\ is shown for an illustrative set of parameters spanning the range of plausible values for $\mu$ and $\sigma_{\rm host}$.}
\end{figure}

\medskip

\begin{figure}[htbp!] 
\includegraphics[width=8cm]{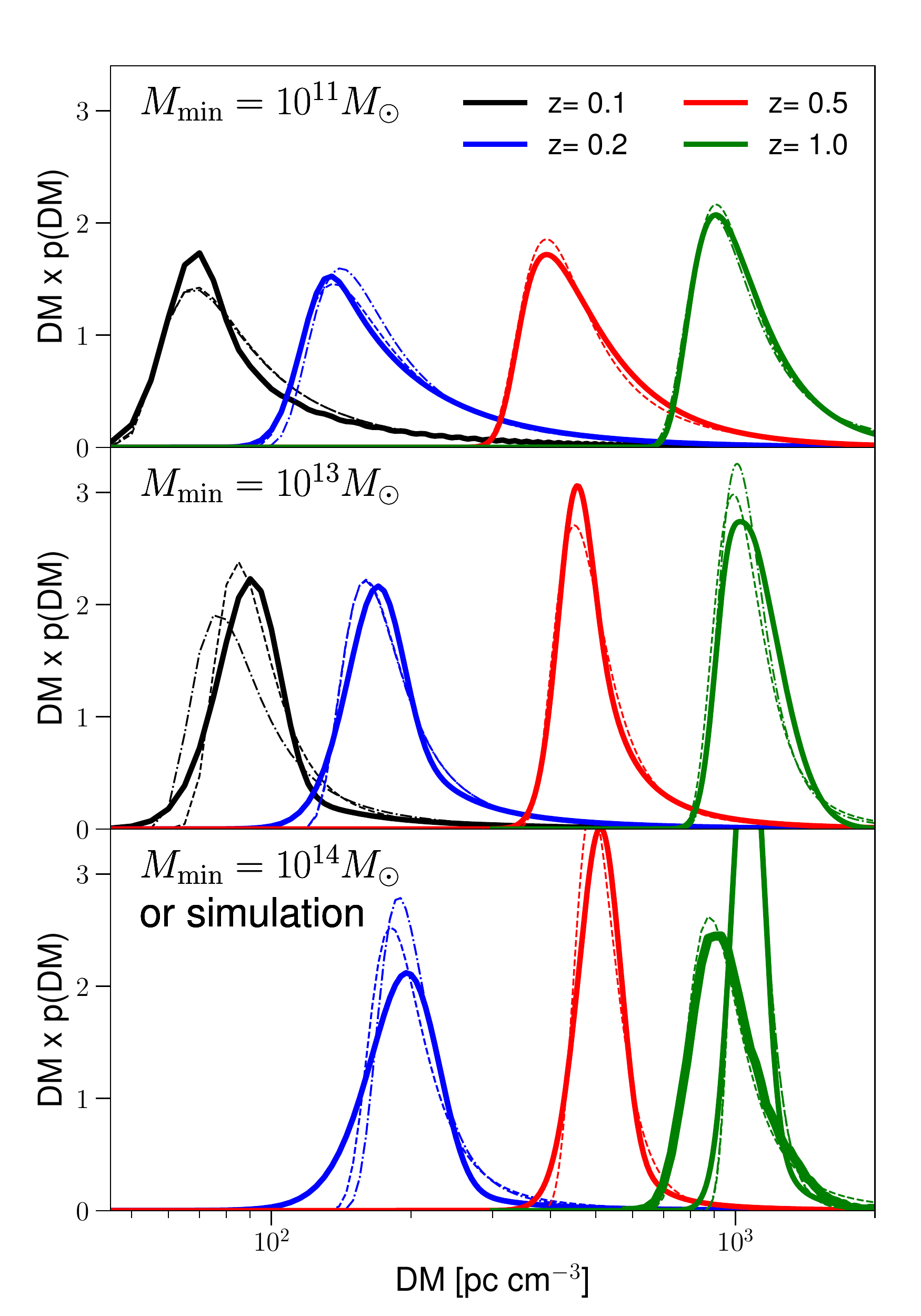}
\caption{
{\bf The expected contribution of the cosmic baryons to the dispersion measure.} The probability distribution of DM$_{\rm cosmic}$ in semi-analytic models and simulations, as encoded in black, blue, red and green in order of increasing redshift, is compared to the analytic form used in our analysis (eq.~\ref{eqn:MHR}).  The thinner solid curves show semi-analytic models$^{11}$
in which the minimum halo mass that can resist feedback and retain its gas is given by $M_{\rm min}$.  The dashed curves are the best-fit analytic function, and the dot-dashed curves assume the $\sigma_{\rm DM} = F z^{-1/2}$ scaling from the $z=0.5$ best-fit for which $F= 0.32, 0.15, 0.09$ for the top, middle and bottom panels respectively.  Because of the success of this Euclidean-space scaling, we adopt it in our analysis.  The thicker green solid curve in the bottom panel is calculated from a hydrodynamic simulation$^{63}$.}
\label{MHRfits}
\end{figure}

\medskip

\begin{figure}[htbp!]
\includegraphics[width=1.0\linewidth]{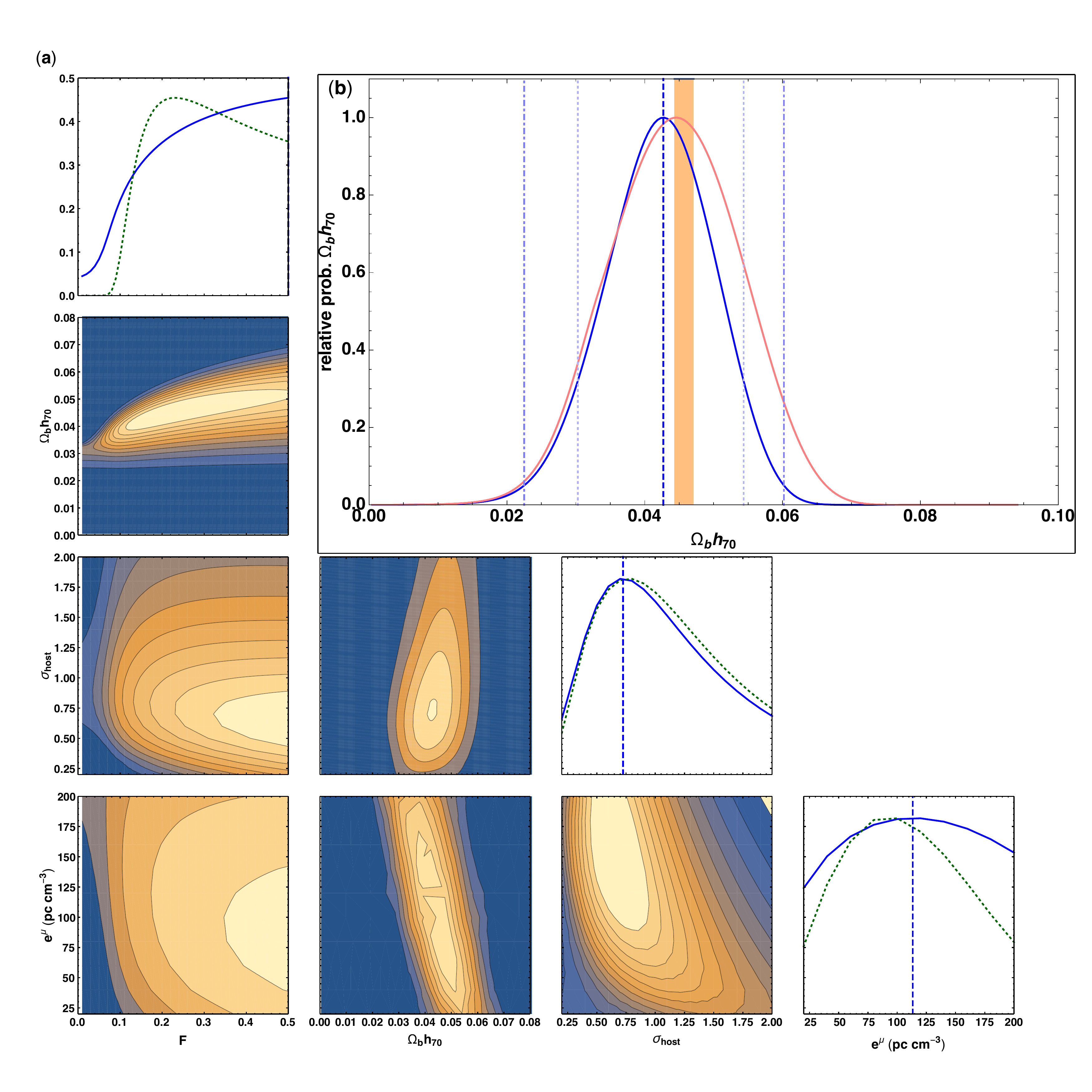} 
    \caption{{\bf The density of cosmic baryons derived from the extended FRB sample.} The constraints on the IGM parameters $\Omega_b h_{70}$ and $F$, and the host galaxy parameters $\mu$ and $\sigma_{\rm host}$ for a log-normal host galaxy DM distribution are shown in an identical manner to Figure 3, but derived using the seven-burst sample (i.e.\,\,including the five gold-standard bursts as well as FRBs\,190523 and 190611).
    \label{fig:sevenburst}} 
\end{figure}

\medskip

\begin{figure}[htbp!]
\includegraphics[width=\linewidth]{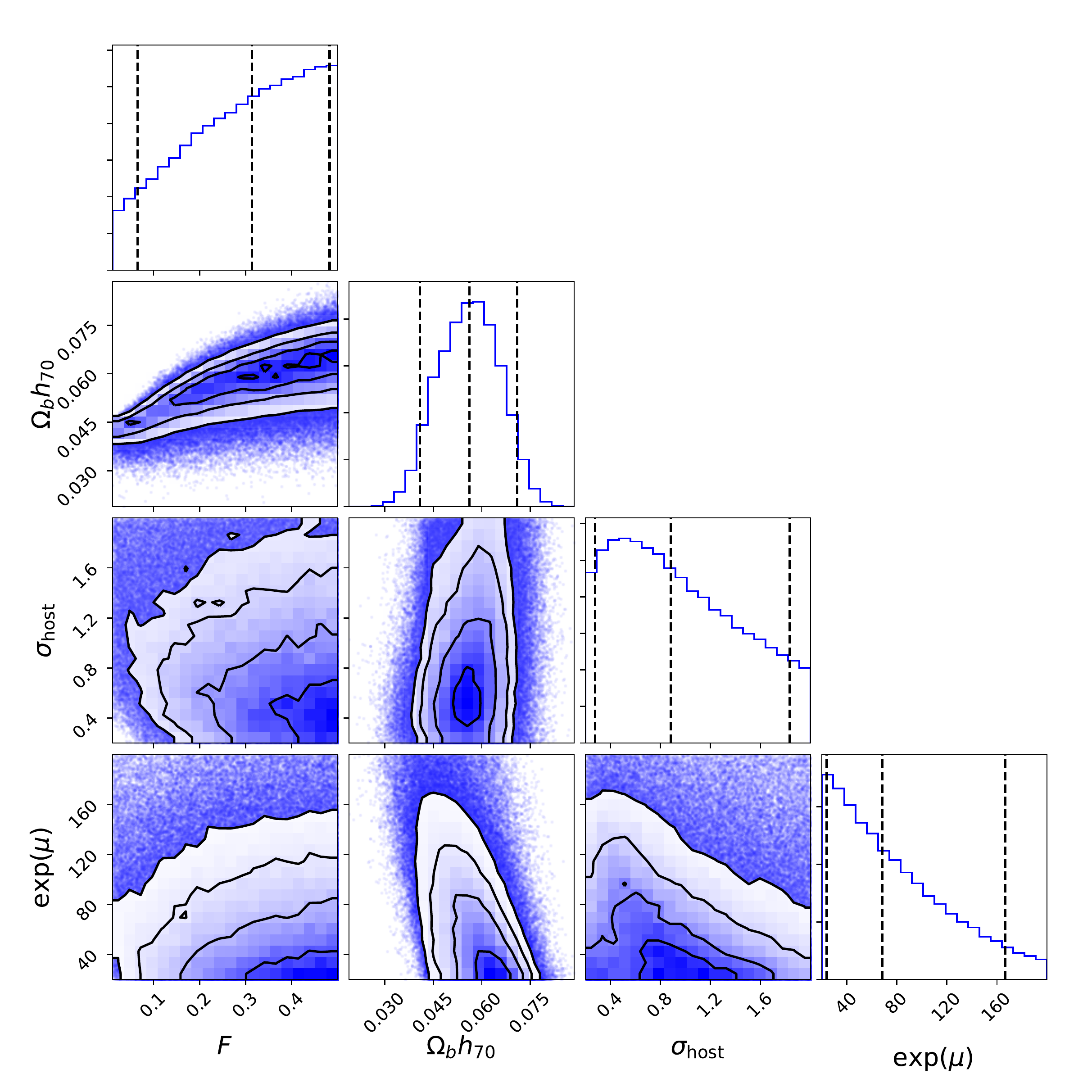}
\caption{{\bf Constraints on the comsic baryon density and FRB host galaxy parameters derived using a Bayesian approach.} The results of a Markov-Chain Monte Carlo (MCMC) analysis based on our five-FRB gold sample presented in the main text demonstrate broad agreement with the results of the frequentist analysis presented in Figure 3. 
}
\label{fig:mcmc_corner}
\end{figure}








\end{document}